\documentclass[prd,aps,floats,floatfix,superscriptaddress,preprintnumbers,
showpacs,eqsecnum,nofootinbib,twocolumn]{revtex4}
\usepackage{amsfonts,amsmath,amssymb,latexsym,array,
theorem,mathrsfs,subfigure,bm,float,color,graphicx}
\usepackage{multirow}
\usepackage{ulem}
\usepackage{here}

\definecolor{red}{rgb}{1,0,0}
\definecolor{blue}{rgb}{0,0,1}
\definecolor{green}{rgb}{0,1,0}

\DeclareMathAlphabet{\mathpzc}{OT1}{pzc}{m}{it}

\newcommand{\ma}[1]{\mbox{$\mathcal{#1}$}}

\newcommand{\calhR}[1]{\raisebox{2ex}{\tiny ({\em h})}\hspace{-0.8em}{\ma R}}
\newcommand{\gsim}{\,\mbox{
\raisebox{-1.ex}{$\stackrel{\textstyle>}{\textstyle\sim}$}}\,}
\newcommand{\lsim}{\,\mbox{
\raisebox{-1.ex}{$\stackrel{\textstyle<}{\textstyle\sim}$}}\,}

\newcommand{\mlv}{m_{\mathrm{LV}}}

\newcommand{\N}{{\rm I\kern-.22em N}}
\newcommand{\Z}{{\sf Z\kern-.42em Z}}
\newcommand{\R}{{\rm I\kern-.22em R}}
\newcommand{\C}{{\rm\kern.22em\rule[.1ex]{.06em}{1.4ex}\kern-.28em C}}
\newcommand{\Q}{{\rm\kern.22em\rule[.1ex]{.06em}{1.4ex}\kern-.28em Q}}
\newcommand{\F}{{\sf F\kern-.42em F}}
\newcommand{\Y}{{\sf Y\kern-.62em Y}}
\newcommand{\X}{{\sf X\kern-.61em X}}
\newcommand{\G}{{\sf G\kern-.61em G}}

\begin{document}

\preprint{WU-AP/1809/18}

\title{
Stable Singularity-free Cosmological Solutions in non-projectable Ho\v{r}ava-Lifshitz Gravity
}


\author{Mitsuhiro {\sc Fukushima}}
\email{m.fukushima"at"aoni.waseda.jp}
\address{Department of Physics, Waseda University, 
Okubo 3-4-1, Shinjuku, Tokyo 169-8555, Japan}

\author{Yosuke {\sc Misonoh}}
\email{y"underscore"misonou"at"moegi.waseda.jp}
\address{Department of Physics, Waseda University, 
Okubo 3-4-1, Shinjuku, Tokyo 169-8555, Japan}

\author{Shoichiro {\sc Miyashita}}
\email{miyashita"at"gravity.phys.waseda.ac.jp}
\address{Department of Physics, Waseda University, 
Okubo 3-4-1, Shinjuku, Tokyo 169-8555, Japan}

\author{Seiga {\sc Sato}}
\email{s.seiga"at"gravity.phys.waseda.ac.jp}
\address{Department of Physics, Waseda University, 
Okubo 3-4-1, Shinjuku, Tokyo 169-8555, Japan}


\date{\today}

\begin{abstract}
We find stable singularity-free cosmological solutions 
in non-flat Friedmann-Lema\^{i}tre-Robertson-Walker (FLRW) spacetime in the context of Ho\v{r}ava-Lifshitz (HL) theory.
Although we encounter the negative squared effective masses of the scalar perturbations in the original HL theory,
the behaviors can be remedied by relaxing the projectability condition.
In our analysis, the effects from the background dynamics are taken into account 
as well as the sign of the coefficients in the quadratic action for perturbations. 
More specifically, we give further classification of the gradient stability/instability into five types.
These types are defined in terms of the effective squared masses of perturbations $\mathcal{M}^2$, 
the effective friction coefficients in perturbation equations $\mathcal{H}$ and
these magnitude relations $|\mathcal{M}^2|/\mathcal{H}^2$.
Furthermore, 
we indicate that oscillating solutions possibly show 
a kind of resonance especially in open FLRW spacetime.
We find that the higher order spatial curvature terms with Lifshitz scaling $z=3$ are significant
to suppress the instabilities due to the background dynamics.
\end{abstract}



\maketitle

\section{Introduction}

From the invention of Big Bang theory, 
resolving the cosmological initial singularity problem has been one of the most intriguing 
research frontier in theoretical physics.
According to the singularity theorem proved by Hawking and Penrose\cite{singularity_theorem}, 
a spacetime singularity must exist at  a finite past of the Big Bang Universe 
based on the general relativity (GR) unless some unnatural conditions are imposed.
Even if the inflation which resolves the fine tuning problems in the very early stage 
of the Universe is introduced, the initial singularity cannot be remedied\cite{inflation_not_complete}.

Since the initial singularity spoils the predictability
at the beginning of the Universe, 
many researchers have been proposed cosmological scenarios to remove 
the singularity. 
For example, in the context of braneworld\cite{braneworld_bounce},
string gas\cite{string_gas}, 
loop quantum gravity\cite{LQG_bounce1,LQG_bounce2,LQG_bounce3},
Horndeski theory (or generalized Galileon)\cite{genesis1,genesis2,genesis3,genesis4, Galileon_bounce,Galileon_bounce2,Galileon_bounce3},
and the non-local gravitational theory\cite{non-local_bounce}.
In spite of these efforts, we have not achieved the complete singularity-free
cosmological scenario, yet.
Recently, striking studies have been conducted.
A quite wide class of singularity-free cosmological solutions 
in Friedmann-Lema\^{i}tre-Robertson-Walker (FLRW)  spacetime
is proved to be unstable\cite{no-go_Horndeski, no-go_multi, no-go_non-flat, no-go_papers}.
Since the no-go theorem is established based on the Horndeski theory, 
in other words, the most generalized scalar-tensor theory 
whose equation of motion is up to second order\cite{Horndeski_theory},
one may consider it is difficult to find stable cosmological solutions
without a singularity.
However, certain loopholes of the no-go theorem are known\cite{bounce_EFT,no-go_loopholes1,
no-go_loopholes2,no-go_loopholes3}.
One example of such loophole is to consider a Lorenz violating gravitational theory. 
The no-go theorem is possibly violated by introducing higher order spatial derivative
terms\cite{bounce_EFT}.

Ho\v{r}ava-Lifshitz (HL) theory is known as a gravitational theory without Lorentz symmetry, 
which is a candidate for quantum gravity\cite{Horava} (the recent progresses of the HL theory are reviewed in \cite{HL-review2}).
As we have known, the spin-2 gravitational field cannot be quantized 
in perturbative approach, 
that is, there is infinite number of loop diagrams with ultraviolet divergence.
In contrast, the HL theory realizes renormalizability 
at least at power-counting level by introducing the Lifshitz scaling, 
\begin{eqnarray}
t\to b^{-z} t~~\mathrm{and}~~x^i \to b^{-1}x^i \,,
\end{eqnarray}
which is an anisotropic scaling between time $t$ and space $x^i$.
A dynamical critical exponent $z$ characterizes a degree of anisotropy.
If $z$ is equal to or greater than the spatial dimension, 
ultraviolet divergence can be suppressed by finite number of counter terms\cite{Horava,ref_renormalize_HL, power-counting_renormalizability}.

In the context of the HL theory, a number of attractive applications 
for cosmology have been conducted (see \cite{HL-review1} for a review): 
for example, primordial perturbations\cite{CP1, CP2, CP3, CP4, CP5}, 
gravitational waves\cite{HLC1, GW-HL} 
and other cosmological aspects\cite{HLC2,DM_integrate, HLC3, dS_projectable}.
It is remarkable that the bouncing and oscillating solutions are discovered
as singularity-free cosmological models in HL theory\cite{bounce_HL1, bounce_HL2, bounce_HL3, bounce_HL4,
previous_oscillating_universe, Bianchi_IX}.
Lifshitz scaling terms up to $z=3$ which realizes
the power-counting renormalizability in four-dimensional spacetime 
derive squared and cubed three-dimensional Ricci curvatures in the action.
These terms behave as radiation-like and stiff matter-like components in non-flat 
FLRW spacetime.
As we will mention, these components can effectively violate the energy condition,  
which is one of the postulates in the singularity theorem.

In our previous paper\cite{previous_stability_bounce}, we investigated the stability of the singularity-free
cosmological solutions based on the {\it projectable} HL theory.
Namely, we impose following the {\it projectability condition}:
 \begin{eqnarray}
N(t, x^i) \to N(t) \,,
 \end{eqnarray}
 where, 
$N$ is the lapse function which is one of the Arnowitt-Deser-Misner (ADM) variables.
It turns out that the HL theory with the projectability condition suffers from the instability 
when the effects of higher order curvature 
terms are irrelevant\cite{problem_projectable1,problem_projectable2,problem_projectable3,problem_projectable4}.
More specifically, we encounter the negative squared effective masses 
of scalar perturbations in FLRW spacetime.
To suppress the instability, we must consider a strong Hubble friction 
by introducing a positive cosmological constant. 

In fact, the pathological behavior in infrared region is possibly remedied by considering a theory without the projectability condition,
that is the {\it non-projectable} HL theory\cite{healthy_extension}.
It is discovered that the gradient instability in flat FLRW spacetime can be avoided\cite{NP_stability1,NP_stability2}.
Therefore it is expected that the infrared instabilities of the singularity-free solutions 
in non-flat FLRW spacetime are improved.

Although the non-projectable HL theory possesses 
the attractive feature\cite{HL-review2,EA,HL_BHTS1,HL_BHTS2,HL_BHTS3,HL_BHTS4, 
HL_BHTS5,HL_BHTS6, inflation_Horava}, 
the most general form of the gravitational action is extremely complicated.
Since what we would like to clarify is to examine whether the non-projectable HL theory can improve
the infrared behavior after singularity avoidance or not, 
we will focus only on an additional term which is dominant in infrared regime, 
which is the {\it minimally extended} HL action.

The rest of this paper is organized as follows.
In Sec.~\ref{minimally_extended_NPHL}, 
we construct the non-projectable HL action with the minimal extension from the projectable one.
The background dynamics in non-flat FLRW spacetime and the classification of the singularity-free solutions are
reviewed in Sec.~\ref{background_dynamics}.
In Sec.~\ref{pert_FLRW_HL}, 
we perform the perturbative analysis by deriving the quadratic action.
The decomposition of the perturbation modes and the manner of the gauge fixing are also summarized.
Then, in Sec.~\ref{stability}, we discuss the stability of singularity-free solutions in non-flat FLRW spacetime.
We give further classification of the gradient stability/instability into five types
to consider the background effects. 
Sec. \ref{conclusion} is devoted to summary and discussions.


\section{minimally extended non-projectable Ho\v{r}ava-Lifshitz theory}
\label{minimally_extended_NPHL} 
As we mentioned, Lifshitz scaling with $z >1$ induces the anisotropy 
between space and time, which means 
the general covariance in four-dimensional spacetime is no longer valid.
We instead find a fundamental symmetry called {\it foliation-preserving diffeomorphism}:
\begin{eqnarray}
t \to t + f(t)\,,~x^i \to x^i + \zeta^i (t,x^j) \,, \label{FPD}
\end{eqnarray}
namely, a boost transformation is prohibited. 
Thus, it is clearly preferable to adopt the ADM quantities, the three-dimensional spatial metric $g_{ij}$,
the lapse function $N$ and the shift vector $N^i$ as the fundamental variables.
In order to preserve the invariance under (\ref{FPD}), the terms in action must be composed of the following quantities: 
The  extrinsic curvature which is defined by
 \begin{eqnarray}
 \mathcal{K}_{ij} := {1 \over 2N} \left( \dot{g}_{ij} - \nabla_i N_j - \nabla_i N_j \right)\,,
 \end{eqnarray}
 where, the {\it dot} symbol represents the derivative with respect to time coordinate and 
 $\nabla_i$ denotes the spatial covariant derivative. 
 The three dimensional Ricci curvature $\mathcal{R}_{ij}$ associated with 
 the spatial covariant derivative. 
 Since we relax the projectability condition, the spatial dependence of the lapse function
 is restored. 
 Thus, the following vector quantity can be included in the action:
  \begin{eqnarray}
 \Phi_i := {\nabla_i N \over N} \,,
 \end{eqnarray} 
 which represents the acceleration of rest observer on three-dimensional hypersurface.
Furthermore, spatial covariant derivatives of these quantities also form 
the invariant action, i.e., $\nabla_i \mathcal{K}_{jk}, \nabla_i \mathcal{R}_{jk}, \nabla_i \Phi_j, \nabla_i \nabla_j \mathcal{K}_{kl}$
 and so on.
 
Although we can include every scalar quantities which are composed of these variables 
to construct the invariant action, our model is restricted as follows:
\begin{eqnarray}
  S = {\mlv^2 \over 2} \int dt \ d^3x \, \, \left( \mathcal{L}_{\mathrm{K}}
+\mathcal{L}_{\mathrm{P}} +\mathcal{L}_{\mathrm{NP}} \right) \,, \label{HL_action}
 \end{eqnarray}
 with
   \begin{eqnarray}
   \mathcal{L}_{\mathrm{K}} &:=& N\sqrt{g}  \left( \mathcal{K}_{ij} \mathcal{K}^{ij} -\lambda \mathcal{K}^2  \right)  
 \,,
   \\  \notag \\
 \mathcal{L}_{\mathrm{P}}&:=& -N \sqrt{g} \Big[ 
 \mathcal{V}_{z=1} + \mlv^{-2} \mathcal{V}_{z=2} + \mlv^{-4} \mathcal{V}_{z=3}
 \Big]
 \,, 
 \label{potential} \\ \notag  \\
 \mathcal{L}_{\mathrm{NP}} &:=& N \sqrt{g} \, \mathcal{V} [\Phi_i] \,,~ \\ \notag \\
 \mathcal{V}_{z=1} &:=& 2\Lambda + g_1\mathcal{R} \,, \notag \\
\mathcal{V}_{z=2} &:=& g_2 \mathcal{R}^2  + g_3 \mathcal{R}^{i}_{~j} \mathcal{R}^{j}_{~i} \,, \notag \\
\mathcal{V}_{z=3} &:=& g_4 \mathcal{R}^3 + g_5 \mathcal{R}\, \mathcal{R}^{i}_{~j} \mathcal{R}^{j}_{~i} 
+g_6 \mathcal{R}^{i}_{~j} \mathcal{R}^{j}_{~k} \mathcal{R}^{k}_{~i} \notag \\
&&+g_7 \mathcal {R} \nabla^2 \mathcal{R} +{g_8 \nabla_{i} \mathcal{R}_{jk} \nabla^{i} \mathcal{R}^{jk}} \,, \label{potential_p}
 \end{eqnarray}
where, $m_{\mathrm{LV}}$ is a Lorentz violating scale which is expected to be around the Planck scale, 
$\lambda$ and $g_i$ ($i=1$-$8$) are dimension-less coupling constants 
and $\Lambda$ is a cosmological constant.
$\mathcal{L}_{\mathrm{NP}}$  is constructed by the terms including $\Phi_i$ field,
which is with no effect to FLRW background.
The reason for the restriction is to ensure the comparability of our previous result based on the SVW action\cite{SVW_paper}.
As we will see in the next section, 
the background dynamics are identical to those of the SVW action under specific condition.

We further impose a restriction on $\mathcal{L}_{\mathrm{NP}}$.
Recall that the purpose of this paper is to remedy the infrared behavior of bouncing solutions
discovered in the projectable HL theory.
It is naturally expected that the terms with the lowest order of spatial derivative
is essential to stabilize the infrared region. 
In fact, such a term is uniquely determined, that is $\Phi^2 := \Phi_i \Phi^i$.
This diagnosis seems not to be inconsequent.
According to the papers \cite{healthy_extension}, the stabilities of  Minkowski and flat FLRW spacetime 
are remedied by the non-projectable HL theory with $\Phi^2$ term.
Therefore, we do minimally extend the theory by adding only $\Phi^2$ term: 
\begin{eqnarray}
\mathcal{V}[\Phi_i] := \varsigma \, \Phi^2 \,.
\end{eqnarray}
with a dimensionless coupling constant $\varsigma$.

In the rest of this paper, the unit $\mlv = 1$ is adopted.
We additionally can set the value of the coupling constant by rescaling the
 time coordinate.
Thus, we set $g_1 = -1$, which is equivalent to take a coordinate in which
the propagating speed of the spin-2 gravitational wave in infrared limit is precisely unity\cite{GW, GW-HL}.

\section{background solutions in FLRW spacetime}
\label{background_dynamics}
The one of the striking differences from the projectable HL theory is the structure of the basic equation system,
that is, the localness of the Hamiltonian constraint is recovered by relaxing the projectability condition.
In this section, we briefly summarize the structure of the basic equation system and
the classification of singularity-free solutions in FLRW spacetime.

\subsection{Basic equations}
To unify the description, we consider the FLRW spacetime in a spatial coordinate system $x^i = (\chi, \theta, \phi)$:
\begin{eqnarray}
ds^2 
&=& g_{ij} dx^i dx^j \notag \\
&=&  a^2 \left[ d \chi ^2 + f(\chi)^2 (d\theta^2 + \sin^2\theta d\phi^2) \right] \,,
\end{eqnarray}
with 
\begin{eqnarray}
f(\chi) := 
\begin{cases}
\chi & \mathrm{for}~K=0 \\
\sin \chi & \mathrm{for}~K=1 \\
\sinh \chi & \mathrm{for}~K=-1 
\end{cases}\,, \label{def_f}
\end{eqnarray}
where, $a$ is the scale factor which depends only on the cosmic time, 
and $K$ is related to the spatial Ricci curvature as $\mathcal{R} = 6 K/ a^2$.
The cases with $K=0, 1$ and $-1$ correspond to the flat, closed and open FLRW spacetime, respectively.
The domain of the coordinate $\chi$ is defined by $0 \leq  \chi < \infty$ for $K=0, -1$ and
$0 \leq  \chi \leq \pi$ for $K=1$.
Furthermore, angular coordinates $\theta$ and $\phi$ take $0 \leq \theta \leq \pi$ 
and $0 \leq \phi \leq 2\pi$, respectively.

Then, the dynamical equation for the scale factor and the Friedmann equation are 
obtained by taking variation of the action with respect to $a$ and $N$:  
\begin{eqnarray}
&&  2 \dot{H} +  3 H^2  = 
  {2 \over 3\lambda -1} \left( \Lambda - {K \over a^2} - {g_r K^2 \over 3a^4} - {g_s K^3 \over a^6}  \right) \,,  \notag \\  \label{a_eom}  \\ \notag \\
&&   H^2  =  {2  \over 3 (3\lambda-1) }\left[ \Lambda -3{K \over a^2} +{g_r K^2 \over a^4} + {g_s K^3 \over a^6} \right] \,, 
\label{H_constraint}
\end{eqnarray}
where, $H:=\dot{a}/a$ is the Hubble parameter, 
$g_r$ and $g_s$ are the linear combinations 
of the coupling constants\footnote{
It should be noted that we adopt the different definitions of $g_r$ and $g_s$ from
our previous paper\cite{previous_stability_bounce} to simplify the perturbed action.
The previous definitions are
\begin{eqnarray}
g_r := 6K^2 (3g_2 +g_3)\,,~ 
g_s := 12 K^3 (9g_4 +3g_5 +g_6) \,, \notag 
\end{eqnarray}
thus, the sign of $g_s$ is flipped in open FLRW spacetime.
}:
\begin{eqnarray}
g_r &:=& 6 (3g_2 +g_3)\,, \\ 
g_s &:=& 12 (9g_4 +3g_5 +g_6) \,.
\end{eqnarray}
We have already fixed the gauge as $N=1$ and $N_i=0$.
In what follows, we assume $\lambda > 1/3$.

As we know, the equation of the scale factor (\ref{a_eom}) can be derived by
taking derivative of  (\ref{H_constraint}) with respect to the cosmic time.
It means the independent equation is only (\ref{H_constraint}). 
However, the situation with the projectability condition is quite different.
Due to the lack of the local lapse function, 
the Hamiltonian constraint is an integration over whole space. 
Therefore, we have to adopt the scale factor equation instead of 
the global Hamiltonian constraint in the projectable case. 
Friedmann-like equation is derived by integration with respect to the cosmic time:
\begin{eqnarray}
H^2  =  {2  \over 3 (3\lambda-1) }\left[ \Lambda -3{K \over a^2} +{g_r K^2 \over a^4} + {g_s K^3 \over a^6} + { \mathcal{C} \over  a^3} \right] \,, 
\end{eqnarray} 
with dust-like term with an integration constant $\mathcal{C}$\cite{DM_integrate}.
Thus, if we consider $\mathcal{C} = 0$ case in the projectable HL theory, 
the same background FLRW solutions are realized\cite{previous_oscillating_universe, previous_stability_bounce}.

\subsection{Singularity-free background solutions}

In order to investigate the background solutions, it is convenient to rewrite 
the Friedmann equation (\ref{H_constraint}) into
\begin{eqnarray}
{1 \over 2}\dot{a}^2 + \mathcal{U}(a) =0\,, \label{scale_factor_potential}
\end{eqnarray} 
with
\begin{eqnarray}
\mathcal{U} (a) = {1 \over 3\lambda-1 } \left[ K -{\Lambda \over 3}a^2  -{g_r K^2 \over 3a^2} -{g_s K^3 \over 3a^4} \right] \,.
\label{bounce_potential}
\end{eqnarray} 
Since the possible ranges for the scale factor are limited within the region in which $\mathcal{U} \leq 0$, 
the background evolution is completely
determined by the coupling constants in $\mathcal{L}_{\mathrm{P}}$.
It is remarkable that the $g_r$ and the $g_s$ terms simulate a radiation component
and a stiff matter component, respectively.
The important point is that these terms can effectively violate the energy condition
if either or both of $g_r K^2$ and $g_sK^3$ are negative, 
which may lead singularity-free solutions even if we do not introduce some exotic matters.

The classification of the possible solutions
is given in Ref.\cite{previous_oscillating_universe}.
In our analysis, we focus on the following two kinds of singularity-free solutions.
One is a {\it bouncing solution} 
denoted by $\mathcal{B}^{[ \mathrm{sgn}(\Lambda); K]}$,
where the function $\mathrm{sgn}(x)$ gives a sign of $x$. 
An initial contracting universe shifts to expanding phase at $a=a_T$
and keep expanding forever.
The other is {\it oscillating solution} 
denoted by $\mathcal{O}^{[ \mathrm{sgn}(\Lambda); K]}$.
A universe shows periodic oscillating behavior, in other words, 
bounces at $a=a_{\mathrm{min}}$ and recollapses\footnote{
In this paper, a word of {\it recollapse} means that the expanding universe shifts to contracting phase.
Thus, a recollapcing point satisfies the conditions $\mathcal{U}=0$ and $\mathcal{U}' >0$.
} at $a=a_{\mathrm{max}}$.
Therefore the scale factor is limited within $0<a_{\mathrm{min}}\leq a \leq a_{\mathrm{max}}<\infty$.
As we will see, the typical size of the oscillating amplitude is expected to be 
the Lorentz violating scale, thus, it is difficult to represent 
the cyclic universe scenario whose maximum scale factor is macroscopic.

\subsubsection{Without cosmological constant}
In our previous paper, we have found that the stable singularity-free solutions require 
a positive cosmological constant based on the projectable HL theory.
Once the projectability condition is relaxed, it is expected
to find stable bouncing solutions without a cosmological constant
as is the case in flat FLRW spacetime. 
Therefore, we summarize the singularity-free cosmological solutions 
without a cosmological constant in this part. 

Since the sign of $\mathcal{U}$ determines the possible ranges for the scale factor, 
it is convenient to derive the roots of the equation $\mathcal{U}=0$:
\begin{eqnarray}
a^{[K]}_{\pm} = \sqrt{ {K \over 6} \left( g_r  \pm \sqrt{g_r^2  + 12g_s } \right)  }\,.
\end{eqnarray} 
The points $a=a^{[K]}_{\pm}$ correspond to the bouncing or recollapsing points
of the universe.
Of course, the corresponding $a^{[K]}_{\pm}$ must be real and positive to find such points.
By examining the forms of the potential $\mathcal{U}$, 
we find the following three types of  singularity-free solutions. 

\begin{figure}[htbp]
\begin{center}
\includegraphics[width=75mm]{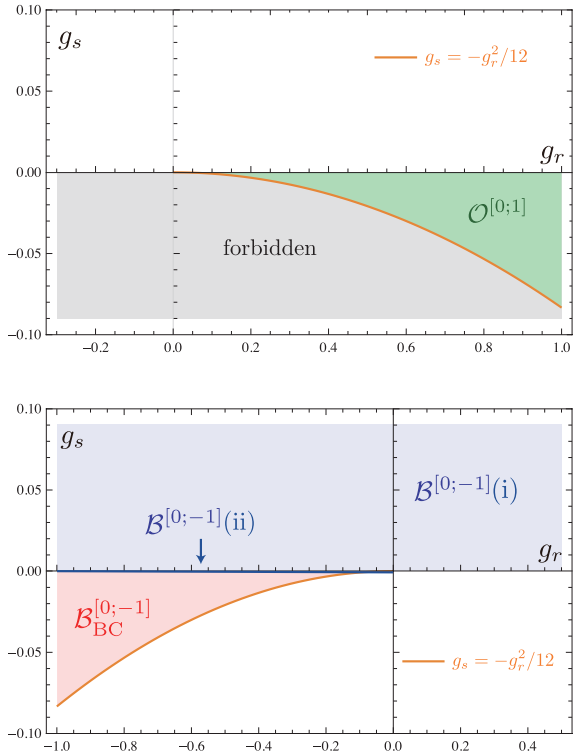}
\end{center}
\caption{
The distribution of the singularity-free solutions for $\Lambda =0$ in $(g_r, g_s)$ plane.
The top and bottom figures correspond to closed and open FLRW spacetime, respectively.
The green, blue and red regions indicate the solutions of $\mathcal{O}^{[0;1]}$, 
$\mathcal{B}^{[0;-1]}$(i) and $\mathcal{B}^{[0;-1]}_{\mathrm{BC}}$, respectively.
$\mathcal{B}^{[0;-1]}$(ii) is distributed on the blue line.
The gray region is forbidden because the Hamiltonian constraint is never satisfied.
}
\label{gr-gs_map}
\end{figure}

\begin{enumerate}
\item[(a)] 
$\mathcal{O}^{[0;1]}$--- 
A universe shows oscillating behavior 
in closed FLRW spacetime ($K=1$), which we call $\mathcal{O}^{[0;1]}$.
This type of the solutions can be found if the following conditions are satisfied.
 \begin{eqnarray}
 &&g_s <0\,,~g_r >0~~~\mathrm{and}~~ g_r^2 +12 g_s >0\notag \\ 
 &&~~~~~~~~~~~~\mathrm{with}~~a_{\mathrm{max}} = a^{[1]}_{+}~~\mathrm{and}~~a_{\mathrm{min}} = a^{[1]}_{-} \,.
 \notag
 \end{eqnarray}
In open FLRW spacetime, this kind of solutions is never found. 

\item[(b)] 
 $\mathcal{B}^{[0;-1]}$--- 
An initially contracting universe shifts to expanding phase at bouncing point $a=a_T$, 
and keeps expanding forever in open FLRW spacetime ($K=-1$).
This type of the solutions is classified into the following two cases: 
\begin{eqnarray}
 (\mathrm{i})&& \forall g_r~~\mathrm{and}~~g_s >0 ~~\mathrm{with}~~a_T = a^{[-1]}_{-}\,. \notag \\ \notag \\
  (\mathrm{ii})&&  g_r <0 ~~\mathrm{and}~~ g_s =0 ~~\mathrm{with}~~a_T = a^{[-1]}_{-} \,. \notag
\end{eqnarray}
The solutions satisfying the conditions (i) and (ii) are referred to as $\mathcal{B}^{[0;-1]}$(i) and
$\mathcal{B}^{[0;-1]}$(ii), respectively.
In closed FLRW spacetime, we cannot find this type of the solutions.
 
 \item[(c)] $\mathcal{B}^{[0;-1]}_{\mathrm{BC}}$---
A universe shows bouncing behavior if the initial value of scale factor 
$a_{\mathrm{ini}}$ is larger than $a_T$,
whereas, falls down into the singularity if $a_{\mathrm{ini}} \leq a_{\mathrm{BC}}$.
Since this type of the solutions possibly induces a Big Crunch (BC),  
we add a subscript BC.
Note that the domain $a_{\mathrm{BC}} < a < a_{T}$ is prohibited 
because the Hamiltonian constraint is never satisfied.
If the following conditions are satisfied in open FLRW spacetime ($K=-1$), 
$\mathcal{B}^{[0;-1]}_{\mathrm{BC}}$ can be found: 
 \begin{eqnarray}
&&g_s <0\,,~g_r <0~~\mathrm{and}~~g_r^2 +12 g_s >0\notag \\ 
 &&~~~~~~~~~~~~~\mathrm{with}~~a_T = a^{[-1]}_{-} ~~\mathrm{and}~~a_{\mathrm{BC}} = a^{[-1]}_{+} \,, \notag 
 \end{eqnarray}
 where, $a_{\mathrm{BC}}$ is a recollapsing point of the Big Crunch universe.  
 This type of bouncing solutions is found only in open FLRW spacetime.
 \end{enumerate}
We show the typical forms of potentials of 
$\mathcal{O}^{[0,1]}$, $\mathcal{B}^{[0,-1]}$ and $\mathcal{B}^{[0,-1]}_{\mathrm{BC}}$
in FIG. \ref{potential}
and the distribution of these types of solutions on $(g_r, g_s)$ plane 
in FIG. \ref{gr-gs_map}.
 \begin{figure}[htbp]
\begin{center}
\includegraphics[width=80mm]{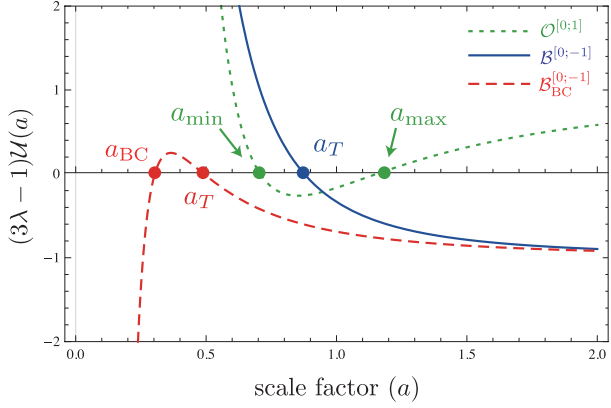}
\end{center}
\caption{
The typical potential forms for singularity-free cosmological solutions without a cosmological constant.
The dotted green, solid blue and dashed red curves indicate 
$\mathcal{O}^{[0;1]}$, $\mathcal{B}^{[0;-1]}$ and $\mathcal{B}^{[0;-1]}_{\mathrm{BC}}$, respectively.
We set the coupling constants to $g_r=11/2$ and $g_s=-2$ for $\mathcal{O}^{[0;1]}$, 
$g_r =-1$ and $g_s =1$ for $\mathcal{B}^{[0;-1]}$,
$g_r =-1$ and $g_s =-1/15$ for $\mathcal{B}^{[0;-1]}_{\mathrm{BC}}$.
}
\label{potential}
\end{figure}

As we will discuss later, 
we can construct the bouncing solutions 
$\mathcal{B}^{[0;-1]}$  and $\mathcal{B}^{[0;-1]}_{\mathrm{BC}}$
whose squared effective masses of scalar perturbations are positive.
Therefore, there is a possibility to construct stable singularity-free solutions without
a cosmological constant. 
It is a striking difference from the projectable HL theory.

\subsubsection{With positive cosmological constant}
Although the properties of singularity-free solutions for $\Lambda >0$ have already given 
in the papers \cite{previous_oscillating_universe, previous_stability_bounce}, 
we again summarize the solutions because the notation is slightly changed from previous ones.
When we consider the case with non-zero cosmological constant, 
it is convenient to introduce the following quantities normalized by a cosmological constant; 
$\tilde{a} := a/\ell$, $\tilde{g}_{r} := g_r/\ell^2$ and $\tilde{g}_s := g_s /\ell^4$ with $ \ell := \sqrt{ 3 / |\Lambda| } $. 
Then, the potential $\mathcal{U}$ is rewritten as
 \begin{eqnarray}
 \tilde{\mathcal{U}} (\tilde{a}) = {1 \over 3\lambda -1} \left[ K - \varepsilon \tilde{a}^2 - { \tilde{g}_r K^2\over 3 \tilde{a}^2} - { \tilde{g}_s K^3\over 3 \tilde{a}^4}  \right] \,,
\end{eqnarray}
where $\varepsilon:=\pm1$ expresses the sign of the cosmological constant.
The three roots of the equation $\tilde{\mathcal{U}} (\tilde{a}) =0$ are given by 
 \begin{eqnarray}
\tilde{a}^{[\varepsilon; K]}_I  &:=& \sqrt{ {K \over 6  \varepsilon} \left[ 2 + {4(1-\tilde{g}_r) \over \tilde{\xi}^{[\varepsilon; K]}_I } 
+  \tilde{\xi}^{[\varepsilon; K]}_I  \right]  }\,,
 \end{eqnarray}
with 
\begin{eqnarray}
\tilde{\xi}^{[\varepsilon; K]}_I &:=& 2^{2/3} (e^{ 2 \pi i / 3 })^I ~\mathrm{pv}\Bigg[ 2 -3\varepsilon \tilde{g}_r  
- 9\tilde{g}_s  \notag \\
&&+9\,\mathrm{sgn}(K)\sqrt{ \left(\tilde{g}_s -\tilde{g}^{[\varepsilon; K]}_{s(+)} \right) 
\left(\tilde{g}_s -\tilde{g}^{[\varepsilon; K]}_{s(-)} \right) } \Bigg]^{1/3}\,, \notag \\
\\
\tilde{g}^{[\varepsilon; K]}_{s(\pm)}  &:=&
{1 \over 9} \left[ 2 -3 \varepsilon \tilde{g}_r  \pm 2 \, \mathrm{sgn}(K) ( 1 - \varepsilon \tilde{g}_r )^{3/2} \right]  \,,
 \end{eqnarray} 
where, $\mathrm{pv}$ means a principal value of cubic root
and $I =1, 2 , 3$. 
 If $\tilde{a}^{[\varepsilon; K]}_I$ takes real and positive value,  
$\tilde{\mathcal{U}} (\tilde{a}) =0$ possesses 
a corresponding root 
in the possible ranges for the scale factor. 
We further 
derive the roots of $\tilde{\mathcal{U}} (\tilde{a}) =0$ with $\tilde{g}_s =0$: 
 \begin{eqnarray}
 \tilde{a}_\pm^{[\varepsilon; K]} := \sqrt{{K \over 2 \varepsilon} 
\left[ 1 \pm \mathrm{sgn}(K) \sqrt{ 1 -{4 \varepsilon \tilde{g}_r  \over 3} } \right] } \,.
 \end{eqnarray} 

\begin{figure}[htbp]
\begin{center}
\includegraphics[width=80mm]{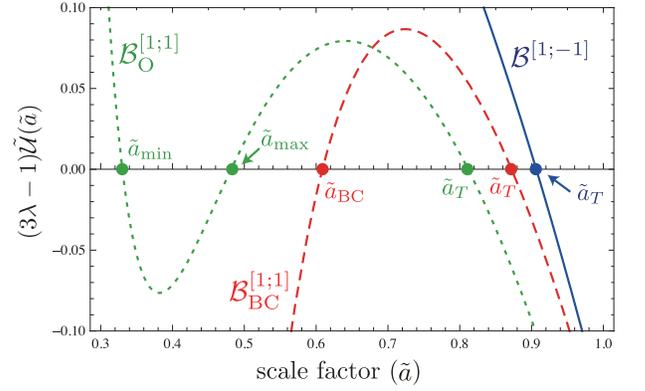}
\end{center}
\caption{
The typical potential forms for singularity-free cosmological solutions with a positive cosmological constant.
The solid blue, dashed red and dotted green curves indicate $\mathcal{B}^{[1;-1]}$, $\mathcal{B}^{[1;1]}_{\mathrm{BC}}$
and $\mathcal{B}^{[1;1]}_{\mathrm{O}}$, respectively.
We set the coupling constants to 
$\tilde{g}_r =1/2$ and $\tilde{g}_s =1/20$ for $\mathcal{B}^{[1;-1]}$, 
$\tilde{g}_r =2/5$ and $\tilde{g}_s =1/9$ for $\mathcal{B}^{[1;1]}_{\mathrm{BC}}$,
$\tilde{g}_r=3/4$ and $\tilde{g}_s=-1/20$ for $\mathcal{B}^{[1;1]}_{\mathrm{O}}$.
}
\label{potential_LambdaP}
\end{figure}

\begin{figure}[htbp]
\begin{center}
\includegraphics[width=75mm]{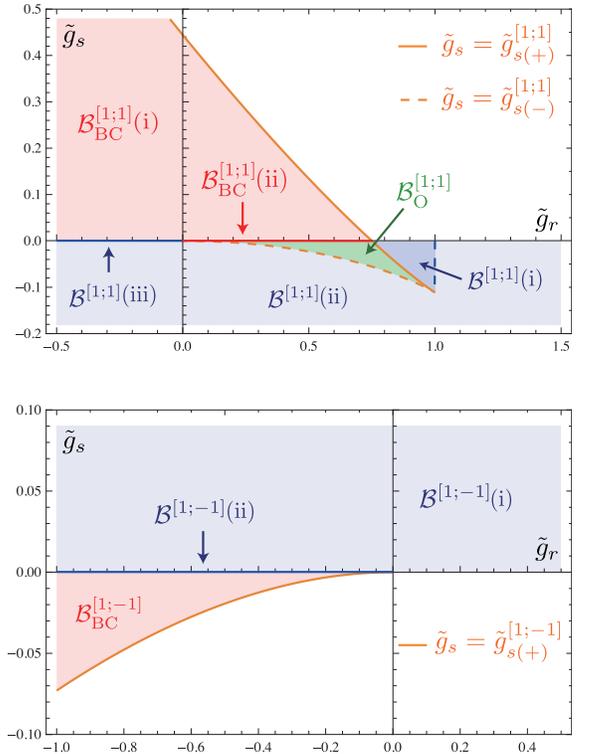}
\end{center}
\caption{
The distribution of the singularity-free solutions for $\Lambda > 0$ in $(g_r, g_s)$ plane.
The top and bottom figures correspond to closed and open FLRW spacetime, respectively.
The blue, red and green regions indicate the solutions of $\mathcal{B}^{[1;K]}$, $\mathcal{B}^{[1;K]}_{\mathrm{BC}}$
and $\mathcal{B}^{[1;1]}_{\mathrm{O}}$, respectively.
$\mathcal{B}^{[1;1]}$(iii), $\mathcal{B}^{[1;-1]}$(ii) and $\mathcal{B}^{[1;1]}_{\mathrm{BC}}$(ii) are
located on the $\tilde{g}_s$ axis.
}
\label{gr-gs_map_LambdaP}
\end{figure}

We classify the singularity-free solutions with a positive cosmological constant 
$(\varepsilon=+1)$ into the following three types: 
\begin{enumerate}
\item[(a)] $\mathcal{B}^{[1;K]}$---
A bouncing solution whose bouncing radius is given as $\tilde{a}_T$.
We refer to this type of the solutions as $\mathcal{B}^{[1;K]}$.
Unlike the case without a cosmological constant, 
the universe after the bounce approaches de Sitter spacetime.
 For closed FLRW spacetime ($K=1$), we can find $\mathcal{B}^{[1;1]}$ as the following three cases: 
\begin{eqnarray}
 (\mathrm{i})&& \tilde{g}_{s(+)}^{[1; 1]}< \tilde{g}_s < 0~~\mathrm{and}~~\tilde{g}_r<1 ~~
  \mathrm{with}~~\tilde{a}_T = \tilde{a}^{[1; 1]}_1\,. \notag \\ \notag \\
 (\mathrm{ii})&&  
\begin{cases}
\tilde{g}_s <0 ~~\mathrm{and}~~\tilde{g}_s< \tilde{g}_{s(-)}^{[1; 1]} & \mathrm{for}~~|2\tilde{g}_r-1|<1\\
\tilde{g}_s <0 & \mathrm{for}~~|2\tilde{g}_r-1| \geq 1
\end{cases} \notag \\
 &&~~~~~~~~~~~~~~~~~~~~~~~~~~~~~~~~~~~~~~
  \mathrm{with}~\tilde{a}_T = \tilde{a}^{[1; 1]}_3\,. \notag \\ \notag \\[-2mm]
(\mathrm{iii})&& \tilde{g}_s=0~~\mathrm{and}~~\tilde{g}_r \leq 0 ~~
 \mathrm{with}~~\tilde{a}_T = \tilde{a}^{[1; 1]}_+\,. \notag 
\end{eqnarray}
The solutions satisfying the conditions (i), (ii) and (iii) are referred to as $\mathcal{B}^{[1;1]}$(i), $\mathcal{B}^{[1;1]}$(ii) and
 $\mathcal{B}^{[1;1]}$(iii), respectively.
 For open FLRW spacetime ($K=-1$), we find $\mathcal{B}^{[1; -1]}$ as the following two cases: 
\begin{eqnarray}
 (\mathrm{i}) && \tilde{g}_s >0~~\mathrm{with}~~\tilde{a}_T = \tilde{a}^{[1; -1]}_3 \,. \notag \\ \notag \\
 (\mathrm{ii}) && \tilde{g}_s =0~~\mathrm{and}~~ \tilde{g}_r <0 ~~\mathrm{with}~~\tilde{a}_T = \tilde{a}^{[1; -1]}_+ \,.  \notag
\end{eqnarray}
$\mathcal{B}^{[1;-1]}$(i) and $\mathcal{B}^{[1;-1]}$(ii) correspond to the conditions (i) and (ii), respectively.

\item[(b)]  $\mathcal{B}^{[1; K]}_{\mathrm{BC}}$---
 A universe with $\tilde{a}_{\mathrm{ini}} \leq \tilde{a}_{\mathrm{BC}}$
 evolves into a Big Crunch, while it shows bouncing behavior 
 if $\tilde{a}_{\mathrm{ini}} \geq  \tilde{a}_{T}  > \tilde{a}_{\mathrm{BC}}$.
 We classify this type of the solutions as $\mathcal{B}^{[1; K]}_{\mathrm{BC}}$.
 Unlike $\mathcal{B}^{[0;K]}_{\mathrm{BC}}$, the asymptotic behavior of spacetime after bounce is de Sitter spacetime.
 For closed FLRW spacetime ($K=1$), we find the following two cases: 
\begin{eqnarray}
 (\mathrm{i})&&0 < \tilde{g}_s < \tilde{g}_{s(+)}^{[1; 1]} \notag \\
 &&~~~~~~~~~~
  \mathrm{with}~~\tilde{a}_{\mathrm{BC}} = \tilde{a}^{[1; 1]}_2~~\mathrm{and}~~\tilde{a}_T = \tilde{a}^{[1; 1]}_3\,. \notag \\ \notag \\
 (\mathrm{ii})&& \tilde{g}_s=0~~\mathrm{and}~~0<\tilde{g}_r< \frac{3}{4} \notag \\
 &&~~~~~~~~~~
  \mathrm{with}~~\tilde{a}_{\mathrm{BC}} = \tilde{a}^{[1; 1]}_- ~~\mathrm{and}~~\tilde{a}_T = \tilde{a}^{[1; 1]}_+\,. \notag
\end{eqnarray}
The solutions satisfying the conditions (i) and (ii) are referred to as $\mathcal{B}^{[1; 1]}_{\mathrm{BC}}$(i) and
 $\mathcal{B}^{[1; 1]}_{\mathrm{BC}}$(ii), respectively.
 For open FLRW spacetime ($K=-1$), the solutions $\mathcal{B}^{[1; -1]}_{\mathrm{BC}}$ can be seen 
when the following conditions are satisfied: 
\begin{eqnarray}
 && 0 > \tilde{g}_s > \tilde{g}_{s(+)}^{[1; -1]}~~\mathrm{and}~~ \tilde{g}_r<0 ~\notag \\
&&~~~~~~~~~~~
 \mathrm{with}~~
 \tilde{a}_{\mathrm{BC}}= \tilde{a}_{2}^{[1; -1]}~~\mathrm{and}~~ \tilde{a}_{T}= \tilde{a}_{3}^{[1; -1]}\,. \notag
\end{eqnarray}
 
\item[(c)] $\mathcal{B}^{[1;1]}_{O}$--- 
A universe in closed FLRW spacetime shows oscillating behavior if the initial scale factor is 
in $\tilde{a}_{\mathrm{min}} \leq  \tilde{a}_{\mathrm{ini}} \leq \tilde{a}_{\mathrm{max}}$, 
or bounces if $\tilde{a}_{\mathrm{ini}} \geq \tilde{a}_T > \tilde{a}_{\mathrm{max}}$.
We classify this type of solutions as $\mathcal{B}^{[1; 1]}_{O}$. 
The subscript represents the oscillating behavior.
This type of the solutions can be found in closed FLRW spacetime ($K=1$)
if the following conditions are satisfied.
\begin{eqnarray}
&& \tilde{g}_s <0 ~~\mathrm{and}~~ \tilde{g}^{[1; 1]}_{s(-)} < \tilde{g}_s < \tilde{g}^{[1; 1]}_{s(+)} \notag \\
&& \mathrm{with}~~ \tilde{a}_{\mathrm{min}} = \tilde{a}^{[1; 1]}_1\,,~ \tilde{a}_{\mathrm{max}} = \tilde{a}^{[1; 1]}_2\
~~\mathrm{and}~~
\tilde{a}_{T} = \tilde{a}^{[1; 1]}_3\,. \notag 
 \end{eqnarray} 
For open FLRW spacetime, we never find this type of solutions.
\end{enumerate}
The typical forms of potentials and the distribution of the singularity-free solutions on $(\tilde{g}_r, \tilde{g}_s)$ plane 
are shown in FIG. \ref{potential_LambdaP} and FIG. \ref{gr-gs_map_LambdaP}, respectively.

\subsubsection{With negative cosmological constant} 
\begin{figure}[htbp]
\begin{center}
\includegraphics[width=80mm]{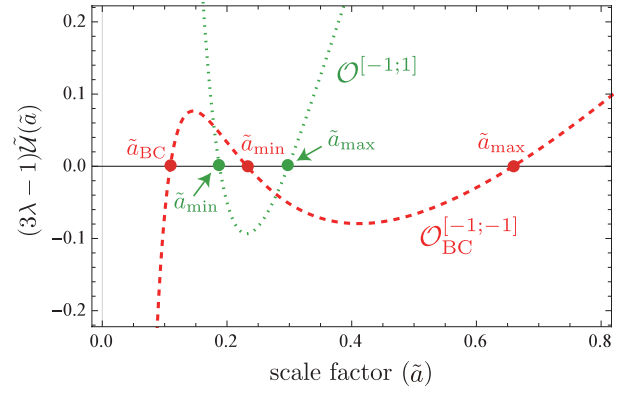}
\end{center}
\caption{
The typical potential forms for singularity-free cosmological solutions with negative cosmological constants.
The dotted green and dashed red curves indicate $\mathcal{O}^{[-1;1]}$ and  
$\mathcal{O}^{[-1;-1]}_{\mathrm{BC}}$, respectively.
We set the coupling constants to $\tilde{g}_r=2$ and $\tilde{g}_s=-1/4$ for $\mathcal{O}^{[-1;1]}$, 
$\tilde{g}_r =-3/4$ and $\tilde{g}_s =-1/20$ for $\mathcal{O}^{[-1;-1]}_{\mathrm{BC}}$.
}
\label{potential_LambdaN}
\end{figure}

\begin{figure}[!htbp]
\begin{center}
\includegraphics[width=75mm]{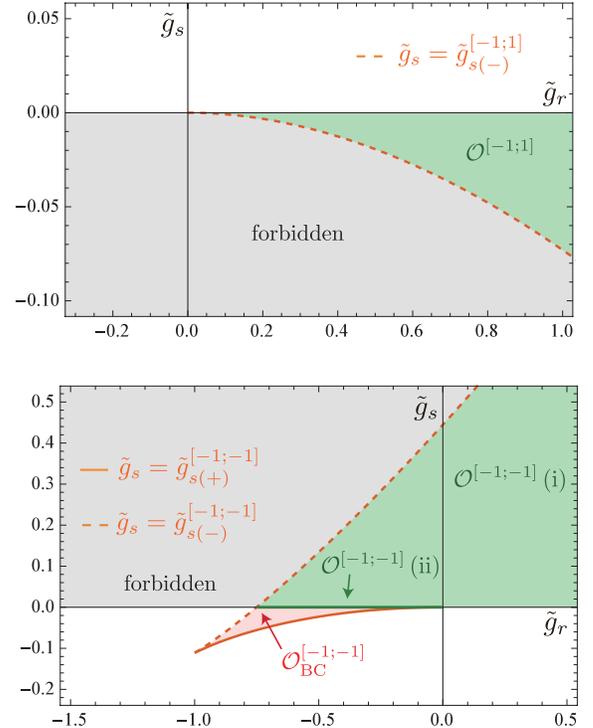}
\end{center}
\caption{
The distribution of the singularity-free solutions for $\Lambda < 0$ in $(g_r, g_s)$ plane.
The top and bottom figures correspond to closed and open FLRW spacetime, respectively.
The green and red regions indicate the solutions of $\mathcal{O}^{[-1;K]}$ and 
$\mathcal{O}^{[-1;-1]}_{\mathrm{BC}}$, respectively.
$\mathcal{O}^{[-1;-1]}$(ii) is located on the $\tilde{g}_s$ axis.
The gray region is forbidden because the Hamiltonian constraint is never satisfied.}
\label{gr-gs_map_LambdaN}
\end{figure}

In this paper, we discuss the stabilities of the oscillating solutions 
as well as the bouncing solutions.
Thus, we do not exclude the solutions with a negative cosmological constant $(\varepsilon=-1)$,
which we cannot construct the bouncing solutions.
We find the following two kinds of singularity-free oscillating solutions:

\begin{enumerate}
 \item[(a)] $\mathcal{O}^{[-1;K]}$--- 
A universe which shows periodic oscillation whose oscillating amplitude is 
given by $\tilde{a}_{\mathrm{min}} \leq \tilde{a} \leq \tilde{a}_{\mathrm{max}}$.
We classify this type of the solutions as $\mathcal{O}^{[-1;K]}$.
If the following conditions are satisfied in closed FLRW spacetime ($K=1$), $\mathcal{O}^{[-1;1]}$ is realized: 
 \begin{eqnarray}
 &&\tilde{g}_{s(-)}^{[-1;1]} <\tilde{g}_s <0 ~~\notag \\
 &&~~~~~~~
\mathrm{with}~~\tilde{a}_{\mathrm{min}}=\tilde{a}_2^{[-1;1]}~~\mathrm{and}~~\tilde{a}_{\mathrm{max}}=\tilde{a}_1^{[-1;1]}\,. \notag \\ \notag
 \end{eqnarray}
In open FLRW spacetime ($K=-1$), we find the two cases of $\mathcal{O}^{[-1;-1]}$ are obtained:
 \begin{eqnarray}
 (\mathrm{i})&& 0< \tilde{g}_s < \tilde{g}_{s(-)}^{[-1; -1]}~~ \notag \\
 &&~
 \mathrm{with}~~\tilde{a}_{\mathrm{min}}=\tilde{a}_2^{[-1;-1]}~~\mathrm{and}~~\tilde{a}_{\mathrm{max}}=\tilde{a}_1^{[-1;-1]}\,. \notag \\[2mm]
 (\mathrm{ii})&& \tilde{g}_s =0~~\mathrm{and}~ -\frac{3}{4}< \tilde{g}_r < 0 \notag \\
 &&~
 \mathrm{with}~~\tilde{a}_{\mathrm{min}}=\tilde{a}_+^{[-1;-1]}~~\mathrm{and}~~\tilde{a}_{\mathrm{max}}=\tilde{a}_-^{[-1;-1]}\,. \notag
 \end{eqnarray}
 The solutions satisfying the conditions (i) and (ii) are referred to as $\mathcal{O}^{[-1;-1]}$(i) and $\mathcal{O}^{[-1;-1]}$(ii), respectively.

 \item[(b)] $\mathcal{O}_{\mathrm{BC}}^{[-1;-1]}$--- 
A universe in open FLRW spacetime shows oscillating behavior if the initial radius of the universe is in
$\tilde{a}_{\mathrm{min}} \leq \tilde{a}_{\mathrm{ini}} \leq \tilde{a}_{\mathrm{max}}$, 
or falls into the singularity if $\tilde{a}_{\mathrm{ini}}<\tilde{a}_{\mathrm{BC}}<\tilde{a}_{\mathrm{min}}$.
We refer to this type of solutions as $\mathcal{O}_{\mathrm{BC}}^{[-1;-1]}$.
For open FLRW spacetime ($K=-1$), we find the solutions under the following conditions:
\begin{eqnarray}
 &&\tilde{g}_s<0\,~~\mathrm{and}~~\tilde{g}_{s(+)}^{[-1;-1]} <\tilde{g}_s <\tilde{g}_{s(-)}^{[-1;-1]} \notag \\
&&~~~~~~~
\mathrm{with}~~\tilde{a}_{\mathrm{BC}}=\tilde{a}_3^{[-1;-1]}\,,~
\tilde{a}_{\mathrm{min}}=\tilde{a}_2^{[-1;-1]} \notag \\
&&~~~~~~~~~~~~~~~\mathrm{and}~~ \tilde{a}_{\mathrm{max}}=\tilde{a}_1^{[-1;-1]}\,. \notag 
 \end{eqnarray}
In closed FLRW spacetime, we cannot construct this type of solutions. 
\end{enumerate}
The typical forms of potentials and the distribution of the singularity-free solutions on $(\tilde{g}_r, \tilde{g}_s)$ plane 
are shown in FIG. \ref{potential_LambdaN} and FIG. \ref{gr-gs_map_LambdaN}, respectively.


\section{perturbation analysis around a non-flat FLRW background}
\label{pert_FLRW_HL}
In this section, we derive the perturbed quadratic action of the minimally extended non-projectable HL theory.
The perturbed ADM variables are defined by 
\begin{eqnarray}
N &=& \bar{N} + \delta N\,,~ \label{pert_ADM_1} \\
N_i &=& \bar{N}_i + \delta N_i\,,~  \label{pert_ADM_2}  \\
g_{ij} &=& \bar{g}_{ij} + \delta g_{ij} \label{pert_ADM_3} \,,
\end{eqnarray}
with
\begin{eqnarray}
\delta N &=& \alpha \,, \notag \\
\delta N_i &=&  \beta_i \,, \notag \\
\delta g_{ij} &=&  h_{ij} +{1 \over 2} \bar{g}^{ab} h_{a i} h_{b j} \,,
\end{eqnarray}
where, $\bar{N}$, $ \bar{N}_i$ and $\bar{g}_{ij}$ denote
the background lapse function, shift vector and three-dimensional induced metric, respectively.
Furthermore, $\alpha$, $\beta_i$ and $h_{ij}$ mean first order 
perturbation of the above ADM variables.
Note that indices of these perturbed variables are raised by $\bar{g}^{ij}$
as $h^i_j:=\bar{g}^{ia}h_{aj}$, $h^{ij}:=\bar{g}^{ia}\bar{g}^{jb}h_{ab}$, 
$h:=h^{i}_i=\bar{g}^{ia}h_{ia}$ and $\beta^i:=\bar{g}^{ia}\beta_a$.

\subsection{Harmonic expansion}
To decompose the perturbations into scalar, vector and tensor modes, 
we perform the harmonic expansion by equipping the set of tensor harmonics: 
\begin{eqnarray}
{\bf Y}&=& \Big\{ {Q}\,, {Q}_i\,, {Q}_{ij}\,, {P}_{ij}\,, {S}_{(o)i}\,, {S}_{(e)i}\,, {S}_{(o)ij}\,, {S}_{(e)ij}\,, 
\notag \\
&&~~ {G}_{(o)ij}\,, {G}_{(e)ij}  \Big\} \,,
\end{eqnarray}
where, $Q\,,~Q_i\,,~Q_{ij}\,,~P_{ij}$ are scalar type harmonics. 
$Q_{ij}$ and $P_{ij}$ are trace and trace-less part, respectively. 
${S}_{(o)i}\,,~{S}_{(e)i}\,,~{S}_{(o)ij}\,,~{S}_{(e)ij}$ are vector type harmonics. 
The symbols ${(o)}$ and ${(e)}$ correspond to the odd parity and the even parity, respectively.
${G}_{(o)ij}\,, {G}_{(e)ij}$ are tensor type harmonics with the odd and the even parity.
The explicit forms and these characteristics can be seen in Appendix A and B in Ref.~\cite{previous_stability_bounce}.

The scalar perturbations of the ADM variables are decomposed into 
\begin{eqnarray}
\alpha^{\mathrm{(scalar)}} &=& \sum_{n,l,m}  \alpha_{(Q)}^{(n;lm)}Q^{(n;lm)}   \label{scalar_N_pert}\,, \\
\beta_i^{\mathrm{(scalar)}} &=& \sum_{n,l,m}  a^2 \left[ \beta^{(n;lm)}_{(Q)} Q^{(n;lm)}_{i} \right] \,, \\
h_{ij}^{\mathrm{(scalar)}} &=& \sum_{n,l,m} a^2 \left[ h^{(n;lm)}_{(Q)} Q^{(n;lm)}_{ij} +h^{(n;lm)}_{(P)} P^{(n;lm)}_{ij}  \right] \,,
\notag \\
\end{eqnarray} 
the vector perturbations also can be expanded by 
\begin{eqnarray}
\beta_i^{\mathrm{(vector)}} &=& \sum_{n,l,m} a^2 \left[ \beta^{(n;lm)}_{(S;o)} S^{(n;lm)}_{(o) i}+\beta^{(n;lm)}_{(S;e)} S^{(n;lm)}_{(e) i} \right] \,, \notag  \\  \\ 
h_{ij}^{\mathrm{(vector)}} &=& \sum_{n,l,m} a^2 \left[ h^{(n;lm)}_{(S;o)} S^{(n;lm)}_{(o)ij} +h^{(n;lm)}_{(S;e)} S^{(n;lm)}_{(e)ij}  \right] \,,\notag \\ 
\end{eqnarray} 
and the tensor part is given by
\begin{eqnarray}
h_{ij}^{\mathrm{(tensor)}} &=& \sum_{n,l,m} a^2 \left[ h^{(n;lm)}_{(G;o)} G^{(n;lm)}_{(o)ij} +h^{(n;lm)}_{(G;e)} G^{(n;lm)}_{(e)ij}  \right] \,. \label{tensor_metric_pert} \notag \\
\end{eqnarray} 
where, the degrees $l, m \in \mathbb{Z}$ are constrained by $0 \leq l \leq n-1$ and $0 \leq |m| \leq l $.
$n \geq 1$ is a continuous real number for $K=0, -1$, 
while a discrete natural number for $K=1$.

Since we relax the projectability condition, the localness of the lapse perturbation $\alpha$
is recovered.
Thus, we expand $\alpha^{(\mathrm{scalar})}$ by the harmonic functions.

\subsection{Gauge fixing}
\label{sec_gauge_fixing}
Before calculating quadratic action, we have to remove the gauge degree of freedom 
from the perturbations.
Since the HL theory respects the {\it foliation-preserving diffeomorphism} (\ref{FPD}), 
the infinitesimal transformations of the perturbed ADM quantities are given by 
\begin{eqnarray}
\alpha^{(\mathrm{gauge})} &=& \partial_t {f}\,,~ \\
\beta_i ^{(\mathrm{gauge})} &=& \partial_t {\zeta}_i -2 H \zeta_i \,, \\
h_{ij}^{(\mathrm{gauge})} &=& 2\nabla_{(i} \zeta_{j)} + 2 H f\, \bar{g}_{ij}  \,.
\end{eqnarray}
We stress that $f$ does not depend on space, thus, 
only $\zeta^i$ can be expanded by the harmonics:
\begin{eqnarray}
\zeta_i &=& \sum_{n,l,m} a^2 \Big[ \zeta^{(n;lm)}_{(Q)} Q_i^{(n;lm)} \notag \\
&&~~~~+ \zeta^{(n;lm)}_{(S;o)} S_{(o)i}^{(n;lm)}  + \zeta^{(n;lm)}_{(S;e)} S_{(e)i}^{(n;lm)}   \Big]\,. 
\end{eqnarray}
Then, the infinitesimal gauge transformations of the harmonic expansion
 (\ref{scalar_N_pert})-(\ref{tensor_metric_pert}) are given by
\begin{eqnarray}
\sum_{n,l,m} \alpha^{(n;lm)}_{(Q)} Q^{(n;lm)} &\to& \sum_{n,l,m} \alpha^{(n;lm)}_{(Q)} Q^{(n;lm)} + \partial_t f \,, 
\label{trans_law_alpha}
\\ \notag \\
\beta^{(n;lm)}_{(Q)} &\to& \beta^{(n;lm)}_{(Q)}+ \partial_t \zeta_{(Q)}^{(n;lm)} \,,  \\ \notag \\
h^{(n;lm)}_{(P)} &\to& h^{(n;lm)}_{(P)} +\sqrt{ {8 (\nu^2 -3K) \over 3} }\zeta_{(Q)}^{(n;lm)}\,, \notag \\
\end{eqnarray}
\begin{eqnarray}
&&\sum_{n,l,m} a^2 h_{(Q)}^{(n;lm)} Q^{(n;lm)}_{ij} \notag \\
&&\to \sum_{n,l,m} a^2 \left[ h^{(n;lm)}_{(Q)}-{2 \nu \over \sqrt{3}}  \zeta_{(Q)}^{(n;lm)}  \right] Q^{(n;lm)}_{ij} \notag 
+2 H f \bar{g}_{ij} \,, \notag \\
\label{trans_law_hQ}
\end{eqnarray}
\begin{eqnarray}
\beta^{(n;lm)}_{(S)} &\to& \beta^{(n;lm)}_{(S)} + \partial_t \zeta_{(S)}^{(n;lm)}  \,,  \\ \notag \\
h^{(n;lm)}_{(S)} &\to& h^{(n;lm)}_{(S)} +\sqrt{2 (\nu^2 -3K) } \zeta_{(S)}^{(n;lm)}\,, 
\end{eqnarray}
\begin{eqnarray}
h^{(n;lm)}_{(G)} &\to& h^{(n;lm)}_{(G)}\,.  
\end{eqnarray}
where, $\nu^2$ is a eigenvalue of the harmonics which is defined by the following regions $(n\geq1)$:
\begin{eqnarray}
&&\nu^2 := 
\begin{cases}
n^2 -1\,,~ n \in \R & \text{for}~ K=0 \\
n^2 -1\,,~ n\in \N &\text{for}~ K=1 \\
n^2 +1\,,~ n \in \R &\text{for}~ K=-1 \\
\end{cases}\,. \label{eigan_nu}
\end{eqnarray}
Since both odd and  even parity modes obey the same transformation law, 
the parity subscripts are abbreviated in vector and tensor perturbations.

We shall eliminate $h^{(n;lm)}_{(P)}$, $h^{(n;lm)}_{(S;o)}$ and $h^{(n;lm)}_{(S;e)}$ by choosing the following gauge 
\begin{eqnarray}
\zeta^{(n;lm)}_{(Q)} &=& - \left[ {8 (\nu^2 -3K) \over 3} \right]^{-1/2} h^{(n;lm)}_{(P)} \,,\\
\zeta^{(n;lm)}_{(S;o)} &=& -\left[2(\nu^2-3K) \right]^{-1/2} h_{(S;o)}^{(n;lm)}  \,,\\
\zeta^{(n;lm)}_{(S;e)} &=& -\left[2(\nu^2-3K) \right]^{-1/2} h_{(S;e)}^{(n;lm)}  \,.
\end{eqnarray}
Unlike the projectable case, we cannot eliminate $\alpha^{(n;lm)}_{(Q)}$.
In what follows, we have abbreviated the superscript $(n;lm)$ 
because the perturbed quantities do not mix with different modes.

\subsection{Quadratic action}
As is the case in the projectable HL theory, the quadratic action can be decomposed 
into the tensor part and the scalar one. 
The vector perturbations are not dynamical.

\subsubsection{Tensor perturbations}
Since the additional perturbation terms coming from $\mathcal{L}_{\mathrm{NP}}$ are only scalar modes, 
the tensor part of the quadratic action is identical to that of the projectable case: 
\begin{eqnarray}
\delta_{(2)} \mathcal{L}^{\mathrm{(tensor)}} 
&=& {a^3 \over 2} \Big[ \mathcal{F}_{(G)} \dot{h}_{(G)}^2 - \mathcal{G}_{(G)} h_{(G)}^2 \Big]\,, \label{action_tensor}
\end{eqnarray} 
where, we introduced $\mathcal{F}_{(G)}$ and $\mathcal{G}_{(G)}$ which can be regarded as the kinetic term and the mass term
of the tensor perturbations, respectively.
These variables, except for the total derivative terms, are given by
 \begin{eqnarray}
\mathcal{F}_{(G)} &:=& 1 \,, \\
\notag \\
\mathcal{G}_{(G)} &:=& {\nu^2 \over a^2}  +{\nu^2 \over 3a^4} \left[ -{2 g_r  K} +3g_3 \nu^2  \right] 
\notag \\
&&
+{\nu^2 \over a^6} \bigg[ 
-{g_s K^2} +6 g_{56} K \nu^2 
+g_8 \nu^2 (\nu^2 -2K)
 \bigg]\,, \notag \\ \label{Gg}
\end{eqnarray}
and we define $g_{56}:=g_5+ g_6$.
The tensor quadratic action is defined for the case with $l \geq 2$, 
because the tensor harmonics $G^{(n;lm)}_{(o)ij}$ and $G^{(n;lm)}_{(e)ij}$ are vanished when $l < 2$.
 
\subsubsection{Scalar perturbations}
The scalar perturbations are drastically changed since the lapse perturbation $\alpha_{(Q)}$ 
cannot be eliminated by the gauge condition.
Furthermore, $\mathcal{L}_{\mathrm{NP}}$ adds a new degree of freedom to scalar perturbations. 
The quadratic action of the scalar perturbations is given by
\begin{widetext}
\begin{eqnarray}
\delta_{(2)} \mathcal{L}^{\mathrm{(scalar)}} &=&
-{a^3 \over 2}(3\lambda-1) \dot{h}_{(Q)}^2 
+\frac{a}{3}(\nu^2-3K) h_{(Q)}^2
-\frac{1}{27a}(\nu^2-3K) \Big[ 2g_r(2\nu^2-3K) + 3g_3\nu^2 \Big]h_{(Q)}^2 \notag \\
&&~~
-\frac{1}{9a^3}(\nu^2-3K) \Big[ g_s K (4\nu^2-9K) +2(3g_{56}-4g_7)K\nu^2 + (-8g_7+3g_8)\nu^2(3\nu^2-10K) \Big]h_{(Q)}^2 
\notag \\
&&~~
+ 2\sqrt{3} a^3 (3\lambda-1)H\dot{h}_{(Q)}\alpha_{(Q)} 
+\frac{4a}{\sqrt{3}}(\nu^2-3K)\Big[ 1 - \frac{2}{3a^2}g_r K - \frac{1}{a^4}K(g_s K -6g_7 \nu^2) \Big] h_{(Q)} \alpha_{(Q)}
\notag \\
&&~~
+2a\Big[ \varsigma \nu^2 -3a^2(3\lambda-1)H^2  \Big] \alpha_{(Q)}^2
-(3\lambda-1)  {2\nu \over \sqrt{3} }  \dot{h}_{(Q)} \beta_{(Q)} 
+4a^3(3\lambda-1)\nu H \alpha_{(Q)}\beta_{(Q)} \notag \\
&&~~
-2a^3\left[ (\lambda-1)\nu^2 +2K  \right] \beta_{(Q)}^2 \,. \label{HL_pert}
\end{eqnarray}
We eliminated $\dot{H}$ by applying the scale factor equation (\ref{a_eom}).  
Since $\alpha_{(Q)}$ and $\beta_{(Q)}$ are not dynamical, 
we can eliminate both of them with the following constraint equations: 
\begin{eqnarray}
\beta_{(Q)} &=& \frac{ (3\lambda-1)\nu}{ (\lambda-1)\nu^2 + 2K }
\Big[ H\alpha_{(Q)} - \frac{1}{2\sqrt{3}}\dot{h}_{(Q)} \Big] \,, \label{alpha_const} \\
\alpha_{(Q)} &=& \frac{(\nu^2-3K)\Bigg[ 
\Big( (\lambda-1)\nu^2 + 2K \Big) \Big( 3a^4 - 2 g_r K a^2 -3K(g_s K -6g_7 \nu^2)  \Big) h_{(Q)}
- 3(3\lambda-1)Ha^6 \dot{h}_{(Q)} \Bigg] } 
{3\sqrt{3}a^4 \Big[ \varsigma \nu^2\Big( (\lambda-1)\nu^2 +2K \Big) +2(3\lambda-1)(\nu^2-3K)H^2 a^2 \Big]}\,, \label{beta_const}
\end{eqnarray}
Substituting, (\ref{alpha_const}) and (\ref{beta_const}) into (\ref{HL_pert}), we achieve the quadratic action of  scalar perturbations:
\begin{eqnarray}
\delta_{(2)} \mathcal{L}^{\mathrm{(scalar)}} &=& {a^3 \over 2} \Big[ \mathcal{F}_{(Q)} \dot{h}_{(Q)}^2 - \mathcal{G}_{(Q)} h_{(Q)}^2 \Big]\,, \label{action_scalar}
\end{eqnarray} 
where,
\begin{eqnarray}
\mathcal{F}_{(Q)} &:=& 
{2\varsigma (3\lambda-1) \nu^2 (\nu^2 -3K) \over 3\varsigma \nu^2 \big[ (\lambda-1)\nu^2  +2K \big] +6(3\lambda-1) (\nu^2 -3K)H^2 a^2  }\,, \label{F_Q} \\ \notag \\ \notag \\
\mathcal{G}_{(Q)} &:=& -{2 \over 3a^2}(\nu^2 -3K)  +{2 \over 27 a^4}(\nu^2 - 3K) \bigg[ {2 g_r }  ( 2\nu^2 - 3K )
+ 3g_3 \nu^2  \bigg] \notag \\
&&+{2 \over 9a^6} (\nu^2 - 3K) \bigg[ {g_s K} (4\nu^2 -9K) +2 (3g_{56} -4 g_7)  K\nu^2 
+(-8 g_7 + 3 g_8) \nu^2 (3\nu^2 - 10K) \bigg] \notag \\
&& +{4 (\nu^2 -3K)^2 \big[ (\lambda-1)\nu^2  +2K \big] \left(  -3g_s K^2 + 18g_7 K \nu^2  -{2} g_r K a^2 + 3a^4   \right)^2 \over 27 a^{10} \Big[ \varsigma \nu^2 \big\{ (\lambda-1)\nu^2  +2K \big\}  +2(3\lambda-1) (\nu^2 -3K)H^2 a^2  \Big] } \notag \\
&& + { 8 K (\nu^2 -3K)^2 \left(  3g_s K^2 +2 g_r K a^2 -3 a^4 \right)\left( 3g_s K^2 -18g_7 K \nu^2  +2 g_r K a^2 -3a^4   \right)
 \over 27 a^{10} \Big[ \varsigma \nu^2 \big\{ (\lambda-1)\nu^2  +2K \big\} +2(3\lambda-1) (\nu^2 -3K)H^2 a^2  \Big] } \notag \\
&& + { 4 (3\lambda-1) (\nu^2 -3K)^2 H^2 
  \left( 3g_s K^2 -18g_7K\nu^2 -2g_r K a^2 +9 a^4 \right)
 \over 9 a^4 \Big[ \varsigma \nu^2 \big\{ (\lambda-1)\nu^2  +2K \big\} +2(3\lambda-1) (\nu^2 -3K)H^2 a^2  \Big] } \notag \\
&& + { 32 (3\lambda-1) (\nu^2 -3K)^3 H^2 \left(  -3g_s K^2 +18g_7 K \nu^2  -2 g_r K a^2 +3a^4   \right) \left(  2g_s K^3 +g_r K^2 a^2  -\Lambda a^6 \right) 
 \over 27 a^8 \Big[ \varsigma \nu^2 \big\{ (\lambda-1)\nu^2  +2K \big\} +2(3\lambda-1) (\nu^2 -3K)H^2 a^2  \Big]^2 }\,. \label{G_Q}
\end{eqnarray} 
\end{widetext}
In closed FLRW spacetime, we focus only on the case with $n \geq 3$.
Actually, the case with $n=1$ corresponds to a constant shift of the scale factor, which is less important, and $n=2$ is not dynamical mode.
Thus, we can regard ($\nu^2 - 3K$) as a positive value.
It should be noted that, even if we take the limit $\varsigma \to 0$,
we cannot replicate the quadratic scalar action in the projectable case
(see (3.26) and (3.27) in ref.\cite{previous_stability_bounce}).
In fact, a difference is caused by the gauge structure.
As we have seen, $\alpha_{(Q)}$ is eliminated by applying constraint equation (\ref{alpha_const}).
Due to the recovering the local lapse function, it cannot be removed by 
global infinitesimal transformation $f$.

\section{stability analysis of the singularity-free solutions}
\label{stability}
To examine whether a singularity-free cosmological solution is truly realized or not, 
it is essential to consider its stability.
If the background solution is unstable with respect to small perturbations, 
the possibility of singularity avoidance may be spoiled.
Thus, we examine the stabilities of singularity-free background solutions which are
shown in previous section. 

When we discuss the stability of a specific solution, 
the sign of $\mathcal{F}$ and $\mathcal{G}$ in the quadratic action provides guideposts
(Since the discussion holds for both scalar and tensor modes, the subscripts are abbreviated.).
The sign of $\mathcal{F}$ is relevant to the ghost instability.
A perturbation mode with negative $\mathcal{F}$ losses the lowest energy state
which leads the fatal collapse of the perturbative approach.
Thus, we consider the condition $\mathcal{F} >0$ as the absolute requirement. 

 {
On the other hand, the case with $\mathcal{G}<0$ is, in general, regarded as the gradient instability.
Such a situation corresponds to the negative squared mass, which seems to induce 
an exponential growth of perturbations.
However, that unstable behavior may be suppressed by the effect of 
background dynamics.
To clarify the effect from the background dynamics, 
we examine the perturbation equation of motions derived by
taking variation of the quadratic action: 
 {
\begin{eqnarray}
\ddot{h} + 3\mathcal{H} \dot{h} + \mathcal{M}^2 h = 0\,,~ 
\label{pert_eom_st}
\end{eqnarray}
where, 
\begin{eqnarray}
&&\mathcal{H} := H  + { \dot{ \mathcal{F} } \over 3 \mathcal{F} }\,,~
\mathcal{M}^2 := { \mathcal{G} \over \mathcal{F} }\,.
\label{def_M2-H}
\end{eqnarray}
From the equation, one can see that the perturbation dynamics is 
influenced  by  {the effective friction coefficient $\mathcal{H}$ as a background effect
as well as the effective squared mass $\mathcal{M}^2$.
The effective friction coefficients of the tensor perturbations 
$\mathcal{H}_{(G)}$ coincide with the Hubble parameter $H$, 
however, those of the scalar perturbations are not: 
\begin{eqnarray}
\mathcal{H}_{(Q)} 
= H \left(1  -{4( \dot{H} + H^2 ) a^2 \over 3\varsigma \nu^2 
\Big[ {(\lambda-1)\nu^2  +2K  \over (3\lambda-1) (\nu^2 -3K) } \Big] +6H^2 a^2  } \right)  \,. 
\end{eqnarray}
We would like to stress that the positive $\mathcal{H}$ generates friction,
on the other hand, the perturbations feel acceleration if $\mathcal{H}$ is negative. 
It is not completely determined by the Hubble parameter $H$ unlike the projectable case.
In what follows, the forces caused by $\mathcal{H}>0$ and $\mathcal{H}<0$ are called as
a {\it $\mathcal{H}$-friction} and a {\it $\mathcal{H}$-acceleration}, respectively.

  Thus, we pay deep attention to the values of $\mathcal{M}^2$ and $\mathcal{H}$ to 
discuss the stability of the perturbations.
In this paper, we classify the stabilities and instabilities into the following five types.
\begin{enumerate}
\item[(i)]
{\it Negative}-$\mathcal{M}^2$ {\it instability}: 
As we mentioned, the negative $\mathcal{M}^2$ causes undesirable exponential 
growth unless it is not suppressed by  {the $\mathcal{H}$-friction}.
Therefore, the unstable behavior due to the negative $\mathcal{M}^2$
is observed in the case with 
(1) $\mathcal{M}^2<0$ and $\mathcal{H}<0$, 
(2) $\mathcal{M}^2<0$ and $|\mathcal{M}^2| \gsim \mathcal{H}^2>0$ 
which means the effect of the third term in LHS of (\ref{pert_eom_st}) is relatively dominant than 
that of the second term. 
If any one of the perturbation modes satisfies this condition, 
we classify the solution is under a {\it negative}-$\mathcal{M}^2$ {\it instability}.
\item[(ii)]
{\it  {$\mathcal{H}$-accelerated instability}}: 
Even if the squared effective mass is positive, 
it is possible to enhance the oscillating amplitude of perturbations.
When  {$\mathcal{H} <0$}, 
the second term in LHS of (\ref{pert_eom_st}) 
reinforces the amplitude rather than friction,  thus  {the perturbation} experiences a $\mathcal{H}$-acceleration. 
Clearly, the accelerating effect is manifested 
if the positive squared effective mass is relatively  {smaller than the effect of $\mathcal{H}$ term}.
Thus, we regard the solution is under a  {$\mathcal{H}$-{\it accelerated instability}} 
if at least one of the perturbation modes satisfies $0 < \mathcal{M}^2 \lsim \mathcal{H}^2$ and $\mathcal{H}<0$.
\item[(iii)]
  { $\mathcal{M}^2$-{\it dominated stability}}:
On the other hand, the negative  {$\mathcal{H}$} is not problematic 
if it is sufficiently suppressed by the heavy positive effective squared mass.
Thus, we call the solution is under a $\mathcal{M}^2$-{\it dominated stability}
if every perturbation mode satisfy $\mathcal{M}^2 \gsim \mathcal{H}^2$ and $\mathcal{H}<0$.
\item[(iv)]
 {\it  {$\mathcal{H}$}-suppressed stability}:
If $\mathcal{H}>0$, we experience a friction effect which suppresses the dynamics of the perturbations.
With this taken into consideration, 
 {we can find a stable solution even if $\mathcal{M}^2 <0$.}
That is, the effective squared mass is negative, however, the unstable behavior is suppressed by the positive $\mathcal{H}$.
We call the solution is under a 
 {\it $\mathcal{H}$-suppressed stability} 
 if every perturbation mode satisfy $|\mathcal{M}^2| \lsim \mathcal{H}^2$ and $\mathcal{M}^2 < 0$.
\item[(v)]
{\it Complete stability}:
Clearly, there is no problematic perturbation dynamics 
if every perturbation mode satisfy $\mathcal{M}^2 > 0$ and $\mathcal{H}>0$.
We call this situation as a {\it complete stability}. 
\end{enumerate}
In TABLE \ref{stability_class_table}, we summarize the classification of the stabilities and instabilities.

In principle, the types of stabilities are entirely determined by the coupling constants in the action.
However, it is difficult to show the explicit conditions for these stabilities
due to the complicated forms of $\mathcal{G}$ and $\mathcal{F}$ for general cases.
Therefore, our main procedure is to numerically trace $\mathcal{G}$ and $\mathcal{F}$
based on the target background solutions.
Before starting numerical analysis, we consider several simplified cases 
in which the stability conditions are given in explicit forms.

\begin{table}[hb!]
\caption{
The classification of stabilities and instabilities which is determined by {$\mathcal{M}^2$ and $\mathcal{H}$}.
}
\begin{center}
\begin{tabular}{ll|c|c|cccccccc}
\hline \hline
\rule[-2mm]{0mm}{5mm}& Stability types &~$\mathcal{H}~$&$~~~ {\mathcal{M}^2}$~~~&~ {$|\mathcal{M}^2| \gsim \mathcal{H}^2 $}& \\
\hline \hline
\rule[-2mm]{0mm}{5mm}&  Negative-$\mathcal{M}^2$ instability~&~$-$~&~$-$~& $\times$
 \\ 
 \rule[-2mm]{0mm}{5mm}& &~$-$~&~$-$~& $\bigcirc$
 \\ 
\rule[-2mm]{0mm}{5mm}& &~$+$~&~$-$~& $\bigcirc$
 \\ 
  \hline 
\rule[-2mm]{0mm}{5mm}&  {$\mathcal{H}$-accelerated instability}~&~$-$~&~$+$~& $\times$
\\ \hline 
\rule[-2mm]{0mm}{5mm}&  {$\mathcal{M}^2$-dominated stability}~ &~$-$~&~$+$~& $\bigcirc$ &\\  \hline 
\rule[-2mm]{0mm}{5mm}&  {$\mathcal{H}$-suppressed stability}~ &~$+$~&~$-$~& $\times$ \\ \hline 
\rule[-2mm]{0mm}{5mm}& Complete stability~  &~$+$~&~$+$~& $\bigcirc$ \\ 
\rule[-2mm]{0mm}{5mm}&  &~$+$~&~$+$~& $\times$ \\ 
\hline \hline 
\end{tabular}
\label{stability_class_table}
\end{center}
\end{table}

\subsection{Late-time universe after bounce}
When the higher order curvature terms are neglected, the forms of $\mathcal{G}$ and $\mathcal{F}$ are quite simplified.
The tensor perturbations are approximated as
\begin{eqnarray}
\mathcal{F}_{(G)} =1\,,~~~ \mathcal{G}_{(G)} \approx { \nu^2 \over a^2 } >0\,,
\end{eqnarray}
which means these are under the complete stability.
On the other hand, those of the scalar perturbations possibly take negative values.
Therefore, we investigate the asymptotic forms of $\mathcal{G}$ and $\mathcal{F}$ in the late-time universe after bounce. 

\subsubsection{Asymptotic flat spacetime}
Although it is impossible to realize a bounce solution in flat spacetime, 
it is expected to be an approximate solution inside the Hubble radius
in the late-time of the universe.
Therefore, we examine the stability of the flat spacetime 
as an asymptotic spacetime after bounce.
We assume $K=0$, $\Lambda =0$ and the scale factor is large, 
then, $\mathcal{F}_{(Q)}$ and $\mathcal{G}_{(Q)}$ are given by\cite{healthy_extension}
\begin{eqnarray}
\mathcal{F}_{(Q)} &=& {2 \varsigma (3\lambda -1)\nu^2 \over 3\varsigma (\lambda -1)\nu^2 + 6(3 \lambda -1) H^2 a^2 }\,, \\
\mathcal{G}_{(Q)} &\approx& {2(2-\varsigma)\nu^2 \over 3\varsigma a^2} \,.
\end{eqnarray} 
Since $\mathcal{G}_{(Q)}$ cannot be positive if $\varsigma <0$, 
we focus  {only} on the case with $\varsigma >0$.
To preserve $\mathcal{F}_{(Q)} >0$ and $\mathcal{G}_{(Q)}>0$,
we must impose 
\begin{eqnarray}
\lambda >1~~\mathrm{and}~~0< \varsigma < 2 \,.
\end{eqnarray} 
Then, the solutions are under the complete stability.

\subsubsection{Asymptotic Milne spacetime}
The Milne universe is also a solution of open FLRW spacetime without a
cosmological constant in deep infrared regime.
The scale factor evolves as
 \begin{eqnarray}
  a \sim \sqrt{ {2 \over 3\lambda-1} } \, t \,.
 \end{eqnarray}
 Then, $\mathcal{F}_{(Q)}$ and $\mathcal{G}_{(Q)}$ are approximated by
\begin{eqnarray}
\mathcal{F}_{(Q)} &\approx& 
\frac{2\varsigma (3\lambda-1) \nu^2 (\nu^2 +3)}{3\big[ \varsigma(\lambda-1)\nu^4 +2(2-\varsigma)\nu^2 +12 \big]} \,, \\ \notag \\
\mathcal{G}_{(Q)}&\approx& { 2\nu^2 (\nu^2 +3) \over 3 a^2 }  \notag \\
&&~\times 
\left[{ (2-\varsigma)(\lambda-1)\nu^2 +2\varsigma +6(\lambda-1) \over \varsigma(\lambda-1)\nu^4 +2(2-\varsigma)\nu^2 +12 }
\right] \,. \notag \\
\end{eqnarray} 
Then, the conditions for the positive $\mathcal{F}_{(Q)}$ and $\mathcal{G}_{(Q)}$
are given by
\begin{eqnarray}
\lambda \geq 1~\mathrm{and}~ 0 < \varsigma < 2 \,.
\label{Milne_stability_cond}
\end{eqnarray} 
The important point is we can stabilize asymptotic spacetime after bounce
without cosmological constant unlike the projectable case.

We also analyze the asymptotic dynamics of the perturbations.
The asymptotic behaviors of the effective friction coefficients are given by
\begin{eqnarray}
\mathcal{H}^2_{(G)} &\sim& {1 \over t^2}\,,~\\
\mathcal{H}^2_{(Q)} &\sim&  {1 \over t^2 }\,,
\end{eqnarray}
and those of the effective squared masses are reduced into 
\begin{eqnarray}
\mathcal{M}^2_{(G)} a^2 &\sim& {( 3\lambda -1 ) \nu^2 \over 2 } \,, \label{MG_Milne} \\
\mathcal{M}^2_{(Q)} a^2 &\sim&  {  (2 - \varsigma)( \lambda - 1 ) \nu^2 + 2\varsigma + 6(\lambda -1)  \over  2 \varsigma } \,.
\label{MQ_Milne}
\end{eqnarray}
Then, the asymptotic dynamics of the perturbations are approximated by
the following form: 
\begin{eqnarray}
h &\sim&  {C_1 \over t } \cos \left[ \sqrt{ \mathcal{M}^2 \, a^2  -1}\,  \ln t   \right] \notag  \\
&& 
+  {C_2 \over t } \sin \left[ \sqrt{ \mathcal{M}^2 \, a^2  -1}\,  \ln t   \right] \,, \end{eqnarray}
with integration constants $C_1$ and $C_2$.
One can see that the perturbation amplitude approaches to zero.

\subsubsection{Asymptotic de Sitter spacetime}
If  a positive cosmological constant is present, 
it may possible that the universe experiences an accelerating expansion 
whose expanding law is given by
 \begin{eqnarray}
 a \sim \exp \left( \sqrt{ {2\Lambda \over 3( 3 \lambda-1 )} } \, t \right) \,.
 \end{eqnarray}

Once the universe enters the accelerating expansion phase, 
the spatial curvature turns to be irrelevant within a few Hubble time, 
then, the spacetime asymptotically approaches to de Sitter spacetime.
In that case, $\mathcal{F}_{(Q)}$ and $\mathcal{G}_{(Q)}$ are approximated by\footnote{
It is natural to consider that
there exists an ultraviolet cut-off momentum $p_{\mathrm{cut}}$ 
at which a non-perturbative quantum effect of gravity cannot be ignored.
More precisely, let $\nu_{\mathrm{cut}}$ be an ultraviolet cut-off mode 
which is related with the cut-off momentum as 
$\nu^2_{\mathrm{cut}}/a_{T}^2 \approx  p^2_{\mathrm{cut}}$
at the early stage of the universe.
Then, we find some finite time so that $\nu^2/a^2 < H^2$ for any $\nu \leq \nu_{\mathrm{cut}}.$
Therefore the all perturbation modes are rapidly suppressed 
by the strong $\mathcal{H}$-friction after that time.
}
\begin{eqnarray}
\mathcal{F}_{(Q)} &\approx& { \varsigma \nu^2 \over 3H^2 a^2} \,, \\
\mathcal{G}_{(Q)} &\approx& \frac{2\nu^2}{27a^4}  
\Bigg[ \frac{27(2+\varsigma)}{4\Lambda}\Big( (\lambda-1)\nu^2+2K \Big) 
\notag \\
&&~~~~~
+ (\nu^2-3K)(4g_r+3g_3) \Bigg]  \,.
\label{asymptotic_dS}
\end{eqnarray}

The positivity of $\mathcal{F}_{(Q)}$ is satisfied if $\varsigma >0$ which is 
consistent with that of in the asymptotic flat spacetime.
We show the explicit conditions for $\mathcal{G}_{(Q)} \geq 0$ in closed 
spacetime: 
\begin{eqnarray}
\displaystyle{3g_3 + 4 g_r + {27 (\lambda -1)(2+\varsigma) \over 4 \Lambda  }} \geq 0\,.
\end{eqnarray}
and in open spacetime: 
\begin{eqnarray}
\displaystyle{3g_3 + 4 g_r + {27 (\lambda -2)(2+\varsigma) \over 10 \Lambda  }} \geq 0\,.
\end{eqnarray}
If the above condition is satisfied, the solutions are under the complete stability 
in the asymptotic de Sitter era.

It is clear that the squared effective masses $\mathcal{M}_{(Q)}^2$
approach asymptotically to zero because it is proportional to $a^{-2}$.
Therefore the case with $\mathcal{G}_{(Q)}<0$ seems not to be quite problematic because 
the solution is under the  {$\mathcal{H}$-suppressed stability}.
To confirm it, we examine the asymptotic dynamics of the perturbations.
 The effective friction coefficients asymptotically  behave as 
\begin{eqnarray}
 \mathcal{H}_{(G)}^2 &\sim& { 2\Lambda \over 3(3\lambda -1)} \,, \\
 \mathcal{H}_{(Q)}^2 &\sim& { 2\Lambda \over 27(3\lambda -1)} \,,
 \end{eqnarray}
 and those of the effective squared masses are
\begin{eqnarray}
\mathcal{M}^2_{(G)} a^2  &\sim& \nu^2 \,, \\
\mathcal{M}^2_{(Q)} a^2 &\sim& \frac{2 H^2 }{9 \varsigma}  
\Bigg[ \frac{27(2+\varsigma)}{4\Lambda}\Big( (\lambda-1)\nu^2+2K \Big) 
\notag \\
&&~~~~~
+ (\nu^2-3K)(4g_r+3g_3) \Bigg]  \,.
 \end{eqnarray}

Substituting them into the perturbation equations of motion, we obtain the asymptotic behaviors.
For the case with $\mathcal{M}^2 >0$, which means the perturbations are under the complete stability,  
the asymptotic dynamics of the tensor perturbations are
\begin{eqnarray}
 h_{(G)} &\sim&  \left[ C_3 \sqrt{ { \mathcal{M}_{(G)}^2 \, a^2 } \over H^2 } e^{- H t}  + C_4  \right] \notag \\
 &&~~~~~~~~~\times \cos \left( \sqrt{ { \mathcal{M}_{(G)}^2 \, a^2 } \over H^2 } e^{- H t} \right) \notag \\
 && - \left[ C_3 - C_4 \sqrt{ { \mathcal{M}_{(G)}^2 \, a^2 } \over H^2 } e^{- H t}  \right] \notag \\
 &&~~~~~~~~~\times \sin \left(  \sqrt{ { \mathcal{M}_{(G)}^2 \, a^2 } \over H^2 } e^{- H t} \right) \,,
\end{eqnarray}
and those of the scalar perturbations are given by
\begin{eqnarray} 
 h_{(Q)} &\sim&  C_5 \cos \left( \sqrt{ { \mathcal{M}_{(Q)}^2 \, a^2 } \over H^2 } e^{- H t} \right) \notag \\
 &&~~~~~ - C_6 \sin \left(  \sqrt{ { \mathcal{M}_{(Q)}^2 \, a^2 } \over H^2 } e^{- H t} \right) \,, 
 \end{eqnarray} 
 with integration constants $C_3$, $C_4$, $C_5$ and $C_6$.
 We find that the perturbations are rapidly suppressed and settled into some constants 
 as $h_{(G)} \sim C_4$ and $h_{(Q)} \sim C_5$.
 Since the constants $C_4$ and $C_5$ are expected to be a typical amplitude 
 of the perturbations when the universe enters the de Sitter era, 
 we generally observe non zero value of perturbation amplitude after entering de Sitter phase.
 
 For the case with $\mathcal{M}^2 <0$,  the dynamics of the tensor perturbations are approximated as
\begin{eqnarray}
 h_{(G)} &\sim&  \left[ C_7 \sqrt{ |{ \mathcal{M}_{(G)}^2 \, a^2 }| \over H^2 } e^{- H t}  + C_8  \right]  \notag \\
 && ~~~~~~\times \cosh \left( \sqrt{ { |\mathcal{M}_{(G)}^2 \, a^2 | } \over H^2 } e^{- H t} \right) \notag \\
 && - \left[ C_7 - C_8 \sqrt{ { | \mathcal{M}_{(G)}^2 \, a^2 | } \over H^2 } e^{- H t}  \right] \notag \\
 && ~~~~~~\times \sinh \left(  \sqrt{ {| \mathcal{M}_{(G)}^2 \, a^2 |} \over H^2 } e^{- H t} \right) \,,
\end{eqnarray}
and those of the scalar perturbations are given by 
 \begin{eqnarray}
 h_{(Q)} &\sim&  C_9 \cosh \left( \sqrt{ { |\mathcal{M}_{(Q)}^2 \, a^2 | } \over H^2 } e^{- H t} \right) \notag \\
 && ~~ -C_{10} \sinh \left(  \sqrt{ {| \mathcal{M}_{(Q)}^2 \, a^2 |} \over H^2 } e^{- H t} \right) \,,
 \end{eqnarray}
 with integration constants $C_7$, $C_8$, $C_9$ and $C_{10}$.
 Then, we also find that the perturbations are rapidly settled into some constant as $h_{(G)} \sim C_8$ and $h_{(Q)} \sim C_9$, 
 which means the $\mathcal{H}$-suppressed stability is truly stabilized state.

\subsection{Whole history of the universe}
Let us turn our attention to the whole history of the universe including bouncing phase.
Although our main analysis depends on the numerical approach, 
we, in advance, consider some simple specific cases again in which explicit conditions can be derived.

\subsubsection{Tensor perturbations}
\label{tensor_stability}
Since the coefficients of kinetic terms in tensor perturbations are unity, 
these are free from the ghost instability.  
Therefore the stabilities are determined only by $\mathcal{G}_{(G)}$.
As we mentioned, the forms of $\mathcal{F}_{(G)}$ and $\mathcal{G}_{(G)}$ in tensor perturbation
are identical to those of the projectable case.
Thus, we summarize the positivity conditions of $\mathcal{G}_{(G)}$, 
which we have argued in previous paper\cite{previous_stability_bounce}.

To investigate the stability of the tensor perturbations, 
we focus on the case with large $\nu^2$.
Then, $\mathcal{G}_{(G)}$ is approximated as 
\begin{eqnarray}
\mathcal{G}_{(G)} \sim { \nu^2 \over a^2 } + g_3 { \nu^4 \over a^4} + g_8 { \nu^6 \over a^6} \,.
\end{eqnarray}
One can see that the ultraviolet stability imposes $g_8 \geq 0$. 

Although the explicit condition for general case is quite complicated, 
we show the conditions with $g_3 = 0$.
In closed spacetime ($K=1$), if one of the following conditions is satisfied, 
$\mathcal{G}_{(G)}$ must be positive for any $a>0$ and $\nu^2 \geq 8$: 
\begin{eqnarray}
(\text{i}) && 0<g_8 \leq -\frac{3g_{56}}{7} \,,~g_r < 0\,,~g_s \leq  -{(g_8-3g_{56})^2 \over g_8} \,; \notag \\
(\text{ii}) && 0<g_8 \leq -\frac{3g_{56}}{7} \,,~g_r \geq 0\,,~ \notag \\
&&~~~~~g_s \leq  -{(g_8-3g_{56})^2 \over g_8} - {g_r^2 \over 9} \,; \notag \\
(\text{iii}) && 0\leq g_8\,,~ -\frac{3g_{56}}{7} < g_8\,,~g_r < 0\,,~g_s \leq 48(g_{56} + g_8) \,; \notag  \\
(\text{iv}) &&  0\leq g_8\,,~ -\frac{3g_{56}}{7} < g_8\,,~g_r \geq 0\,,~ \notag \\
&&~~~~~g_s \leq 48(g_{56} + g_8) - {g_r^2 \over 9} \,; \notag
\end{eqnarray} 
Similarly, in open spacetime ($K=-1$): 
\begin{eqnarray}
(\text{i}) && 0 < g_8 \leq g_{56}\,,~g_r >0  \,,~g_s \leq -{(g_8-3g_{56})^2 \over g_8}\,; \notag \\
(\text{ii}) && 0 < g_8 \leq g_{56}\,,~g_r \leq 0 \,,~g_s \leq -{(g_8-3g_{56})^2 \over g_8} -{g_r^2 \over 9} \,; \notag \\
(\text{iii}) && 0\leq g_8\,,~g_{56} < g_8\,,~g_r >0 \,,~g_s \leq 4(2g_{8} -3g_{56}) \,; \notag \\
(\text{iv}) && 0\leq g_8\,,~g_{56} < g_8\,,~g_r \leq 0 \,,~g_s \leq 4(2g_{8} -3g_{56}) - {g_r^2 \over 9} \,, \notag 
\end{eqnarray}
which are equivalent to $\mathcal{G}_{(G)} \geq 0$ for any $a>0$ and $\nu^2 \geq 2$. 
It is worth mentioning that these stability conditions in open FLRW spacetime are not completely 
conflict, however slightly difficult to be compatible with the bouncing conditions
(see \cite{previous_stability_bounce}).

\subsubsection{Scalar perturbations}
The ghost-free condition for scalar perturbations can be analytically discussed.
 {Let the coupling constants be} $\lambda >1/3$ and $\varsigma >0$, and  {adopting} $\nu^2 -3K >0$,
the condition $\mathcal{F}_{(Q)} >0$ is equivalent to 
\begin{eqnarray}
{\varsigma \nu^2 \left[ (\lambda-1)\nu^2 +2K \right] \over 2(3\lambda-1)(\nu^2 -3K)a^2 } > -H^2 \,.
\end{eqnarray} 
The most stringent conditions are imposed at the bouncing and the recollapse points, i.e., $H=0$:
\begin{eqnarray}
\varsigma >0~\mathrm{and}~
\begin{cases}
\lambda > 1 ~\mathrm{for}~K=0 \\
\lambda \geq 1 ~\mathrm{for}~K=1 \\
\lambda > 2 ~\mathrm{for}~K=-1
\end{cases} \,. \label{ghost_free_cond}
\end{eqnarray}
If the above conditions are satisfied, the ghost instability can be eliminated 
throughout the evolution of the universe.

On the other hand, the positivity conditions of $\mathcal{G}_{(Q)}$ 
cannot be expressed without any approximation due to the extremely complicated form.
Therefore, we restrict our analysis to the case with large $\nu^2$ modes at this stage. 
Then, $\mathcal{G}_{(Q)}$ is given by
\begin{eqnarray}
\mathcal{G}_{(Q)} \approx \Bigg[ (-8g_7+3g_8) + \frac{72g_7^2K^2}{\varsigma a^{4}} \Bigg] \frac{2\nu^6}{3a^6} 
\,,
\end{eqnarray}
Since $\varsigma$ should be positive,  we assume 
\begin{eqnarray}
3g_8\geq8g_7 \,.
\end{eqnarray}
Then, $\mathcal{G}_{(Q)} \geq 0$ for large $\nu$ is guaranteed.

\subsubsection{Numerical analysis}
\label{section_numerical}

\begin{table*}[thb!]
\caption{
 The examples for the stable singularity-free solutions, which mean that these solutions are under the 
 {$\mathcal{M}^2$}-dominated 
 stability  {when $\mathcal{H}<0$} and under the complete stability  {when $\mathcal{H}>0$}.
The types of the solutions are introduced in Sec.~\ref{background_dynamics}.
N/A means that there is no corresponding variables.
}
{\scriptsize
\begin{center}
\begin{tabular}{cc|c||c|c|c|c|c|c|c|c|c|c||c|c||c|c|c|cc}
\hline \hline
\rule[-1mm]{0mm}{4mm}&&~Type~~&~$\Lambda$~~&~$\lambda$~~&~$g_2$~&~$g_3$~~&~$g_4$~&~$g_5$~&~$g_6$~&~~$g_7$~&
~~$g_8$~&~~$\varsigma$~&~~$g_r$~&~~$g_s$~&~~$a_{\mathrm{BC}}$~&~~$a_{\mathrm{min}}$~&~~$a_{\mathrm{max}}$~&~~$a_T$~&\\
\hline \hline 
\rule[-3.5mm]{0mm}{9mm}& (i)~~&$\mathcal{O}^{[0;1]}$~&
~~$0$~~&~~$1$~~&~~$\displaystyle{-{1 \over 18}}$~~&~~$\displaystyle{{1 \over 3}}$~~&~~$\displaystyle{-{1 \over 108}}$~~&~~$0$~~&
~~$\displaystyle{{7 \over 90}}$~~&~~$\displaystyle{{1 \over 2}}$~~&~~$1$~~&~~$1$~~&~~$1$~~&~~$\displaystyle{-{1 \over 15}}$~~&~~N/A~~&~~0.304~~&~~0.491~~&~~N/A~~&
 \\ \hline
\rule[-3.5mm]{0mm}{9mm}& (ii)~~&$\mathcal{B}^{[0;-1]}$~&
~~$0$~~&~~$6$~~&~~$\displaystyle{-{1 \over 18}}$~~&~~$\displaystyle{{1 \over 8}}$~~&~~$0$~~&~~$\displaystyle{{1 \over 3}}$~~&
~~$\displaystyle{-{2 \over 3}}$~~&~~$-1$~~&~~$3$~~&~~$\displaystyle{{3 \over 2}}$~~&~~$\displaystyle{-{1 \over 4}}$~~&~~$4$~~&~~N/A~~&~~N/A~~&~~N/A~~&~~$1.094$~~&
 \\ \hline
\rule[-3.5mm]{0mm}{9mm}& (iii)~~&$\mathcal{B}_{\mathrm{BC}}^{[0;-1]}$~&
~~$0$~~&~~$\displaystyle{5\over2}$~~&~~$\displaystyle{-{1 \over 18}}$~~&~~$\displaystyle{-{1 \over 3}}$~~&~~$\displaystyle{{1 \over 20}}$~~&~~$\displaystyle{-{1 \over 4}}$~~&
~~$\displaystyle{{1 \over 4}}$~~&~~$0$~~&~~$1$~~&~~$\displaystyle{{1 \over 2}}$~~&~~$-3$~~&~~$\displaystyle{-{3 \over 5}}$~~&~~$0.526$~~&~~N/A~~&~~N/A~~&~~$0.851$~~&
 \\ \hline
\rule[-3.5mm]{0mm}{9mm}& (iv)~~&$\mathcal{B}^{[1;1]}$~&
~~$1$~~&~~$1$~~&~~$\displaystyle{{1 \over 6}}$~~&~~$\displaystyle{-{1 \over 6}}$~~&~~$\displaystyle{-{1 \over 12}}$~~&~~$\displaystyle{{1 \over 3}}$~~&
~~$\displaystyle{-{1 \over 3}}$~~&~~$\displaystyle{-{1 \over 2}}$~~&~~$1$~~&~~$\displaystyle{{3 \over 2}}$~~&~~$2$~~&~~$-1$~~&~~N/A~~&~~N/A~~&~~N/A~~&~~$1.525$~~&
 \\ \hline
\rule[-3.5mm]{0mm}{9mm}& (v)~~&$\mathcal{B}_{\mathrm{BC}}^{[1;1]}$~&
~~$2$~~&~~$2$~~&~~$\displaystyle{{1 \over 9}}$~~&~~$\displaystyle{-{1 \over 2}}$~~&~~$\displaystyle{-{1 \over 12}}$~~&~~$\displaystyle{{1 \over 3}}$~~&
~~$\displaystyle{-{1 \over 6}}$~~&~~$\displaystyle{-{1 \over 2}}$~~&~~$1$~~&~~$1$~~&~~$-1$~~&~~$1$~~&~~$0.707$~~&~~N/A~~&~~N/A~~&~~$1.272$~~&
 \\ \hline
\rule[-3.5mm]{0mm}{9mm}& (vi)~~&$\mathcal{B}_{\mathrm{O}}^{[1;1]}$~&
~~$2$~~&~~$1$~~&~~$\displaystyle{{2 \over 45}}$~~&~~$0$~~&~~$\displaystyle{-{1 \over 108}}$~~&~~$\displaystyle{{2 \over 75}}$~~&
~~$0$~~&~~$0$~~&~~$2$~~&~~$\displaystyle{{1 \over 2}}$~~&
~~$\displaystyle{{4 \over 5}}$~~&~~$\displaystyle{-{1 \over 25}}$~~&~~N/A~~&~~$0.256$~~&~~$0.511$~~&~~$1.083$~~&
 \\ \hline
\rule[-3.5mm]{0mm}{9mm}& (vii)~~&$\mathcal{B}^{[1;-1]}$~&
~~$1$~~&~~$\displaystyle{{5 \over 2}}$~~&~~$0$~~&~~$\displaystyle{-{1 \over 6}}$~~&~~$0$~~&~~$\displaystyle{{1 \over 6}}$~~&
~~$\displaystyle{-{1 \over 3}}$~~&~~$\displaystyle{{1 \over 2}}$~~&~~$2$~~&~~$1$~~&~~$-1$~~&~~$2$~~&~~N/A~~&~~N/A~~&~~N/A~~&~~$0.928$~~&
\\ \hline
\rule[-3.5mm]{0mm}{9mm}& (viii)~~&$\mathcal{B}_{\mathrm{BC}}^{[1;-1]}$~&
~~$2$~~&~~$3$~~&~~$\displaystyle{-{1 \over 3}}$~~&~~$\displaystyle{{1 \over 2}}$~~&~~$\displaystyle{-{1 \over 30}}$~~&~~$\displaystyle{{1 \over 5}}$~~&
~~$\displaystyle{-{1 \over 3}}$~~&~~$-1$~~&~~$1$~~&~~$\displaystyle{{3 \over 2}}$~~&~~$-3$~~&~~$\displaystyle{-{2 \over 5}}$~~&~~$0.403$~~&~~N/A~~&~~N/A~~&~~$0.745$~~&
 \\ \hline
\rule[-3.5mm]{0mm}{9mm}& (ix)~~&$\mathcal{O}^{[-1;1]}$~&
~~$-1$~~&~~$1$~~&~~$\displaystyle{-{1 \over 12}}$~~&~~$\displaystyle{{1 \over 2}}$~~&~~$\displaystyle{-{1 \over 108}}$~~&~~$\displaystyle{{1 \over 40}}$~~&
~~$0$~~&~~$1$~~&~~$1$~~&~~$1$~~&~~$\displaystyle{{3 \over 2}}$~~&~~$\displaystyle{-{1 \over 10}}$~~
&~~N/A~~&~~0.282~~&~~0.604~~&~~N/A~~&
 \\ \hline
\rule[-3.5mm]{0mm}{9mm}& (x)~~&$\mathcal{O}^{[-1;-1]}$~&
~~$-1$~~&~~$3$~~&~~$\displaystyle{{1 \over 4}}$~~&~~$\displaystyle{-{1 \over 2}}$~~&~~$0$~~&~~$\displaystyle{-{1 \over 40}}$~~&
~~$\displaystyle{{1 \over 5}}$~~&~~$0$~~&~~$3$~~&~~$1$~~&~~$\displaystyle{{3 \over 2}}$~~&~~$\displaystyle{{3 \over 2}}$~~
&~~N/A~~&~~0.730~~&~~1.821~~&~~N/A~~&
 \\ \hline
\rule[-3.5mm]{0mm}{9mm}& (xi)~~&$\mathcal{O}_{\mathrm{BC}}^{[-1;-1]}$~&
~~$\displaystyle{-{3 \over 2}}$~~&~~$\displaystyle{{5 \over 2}}$~~&~~$\displaystyle{{1 \over 8}}$~~&~~$\displaystyle{-{1 \over 2}}$~~&~~$\displaystyle{{1 \over 216}}$~~&~~$\displaystyle{-{1 \over 60}}$~~&
~~$\displaystyle{{1 \over 150}}$~~&~~$-1$~~&~~$2$~~&~~$\displaystyle{{1 \over 2}}$~~&~~$\displaystyle{-{3 \over 4}}$~~&~~$\displaystyle{-{1 \over 50}}$~~
&~~0.174~~&~~0.507~~&~~1.309~~&~~N/A~~&
 \\ 
\hline \hline 
\end{tabular}
\label{table_stable_sol}
\end{center}}
\end{table*}
 
Before performing the numerical analysis, we summarize the stability conditions 
for each cases.
\begin{enumerate}
\item[(1)] The cases with $K=1$ and $\Lambda \leq 0$: 
Since there exists a finite maximum value of scale factor $a_{\mathrm{max}}$,
i.e., the solutions $\mathcal{O}^{[0; 1]}$ and $\mathcal{O}^{[-1; 1]}$, 
we assume the stability conditions around the bouncing and the recollapsing points  
and in the large $\nu^2$ region, thus, 
\begin{eqnarray}
\varsigma >0\,,~\lambda \geq 1\,,~ g_8\geq 0~\mathrm{and}~3g_8 \geq 8g_7 \,. \label{cond_K1LnP}
\end{eqnarray}  
\item[(2)] The cases with $K=1$ and $\Lambda >0$:
In this cases, we observe $\mathcal{B}^{[1;1]}$, $\mathcal{B}^{[1;1]}_{\mathrm{BC}}$
 and $\mathcal{B}^{[1;1]}_{\mathrm{O}}$ as the singularity-free solutions.  
To guarantee the stabilities around bouncing point (and the recollapsing point if $\mathcal{B}^{[1;1]}_{\mathrm{O}}$), 
in asymptotic de Sitter spacetime 
and in the large $\nu^2$ region, we assume
\begin{eqnarray}
&&\varsigma >0\,,~\lambda \geq 1\,,~
3g_3 +4g_r +{ 27 (\lambda -1) (2+\varsigma)\over 4 \Lambda } \geq 0\,, \notag \\
&&{g_8\geq0~}\mathrm{and}~3g_8 \geq 8g_7 \,.
\end{eqnarray}
If we focus only on the oscillating phase which is observed in the solution $\mathcal{B}^{[1;1]}_{\mathrm{O}}$, 
the stability conditions in the asymptotic de Sitter spacetime  
can be relaxed. Then, we impose the same conditions as
the cases with $K=1$ and $\Lambda \leq 0$, that is, (\ref{cond_K1LnP}).
\item[(3)] The cases with $K=-1$ and $\Lambda = 0$: 
We find $\mathcal{B}^{[0,-1]}$ and $\mathcal{B}^{[0,-1]}_{\mathrm{BC}}$
as the singularity-free solutions whose asymptotic behaviors are the Milne spacetime.
Thus, we impose 
\begin{eqnarray}
0 < \varsigma <2\,,~\lambda > 2\,,~g_8\geq0~\mathrm{and}~3g_8 \geq 7 g_7\,.
\end{eqnarray}
\item[(4)] The cases with $K=-1$ and $\Lambda > 0$: 
In this case, the solutions $\mathcal{B}^{[1,-1]}$ and $\mathcal{B}^{[1,-1]}_{\mathrm{BC}}$
are realized.
Since both of the solutions approaches the de Sitter spacetime after the bounce, 
we impose the conditions as follows:  
 \begin{eqnarray}
&&\varsigma >0\,,~\lambda > 2\,,~
3g_3 +4g_r +{ 27 (\lambda -2) (2+\varsigma)\over 10 \Lambda } \geq 0\,, \notag \\
&&g_8\geq0~\mathrm{and}~3g_8 \geq 8g_7 \,.
\end{eqnarray}
\item[(5)] The cases with $K=-1$ and $\Lambda < 0$: 
The possible singularity-free solutions are $\mathcal{O}^{[-1; -1]}$ and $\mathcal{O}^{[-1; -1]}_{\mathrm{BC}}$.
Therefore we impose the stability conditions around the bouncing and recollapsing points and 
in the large $\nu^2$ region:
 \begin{eqnarray}
&&\varsigma >0\,,~\lambda > 2\,,~g_8\geq0~\mathrm{and}~3g_8 \geq 8g_7 \,. \label{cond_K-1LnP}
\end{eqnarray}
\end{enumerate}

Then, we shall investigate the spacetime stabilities in each cases: 
bouncing solutions with asymptotic de Sitter spacetime, 
bouncing solutions with asymptotic Milne spacetime and
oscillating solutions.
In our numerical analysis, we have found stable  {singularity-free} solutions 
throughout the whole evolutions, which means 
these solutions are under the  {$\mathcal{M}^2$}-dominated stability  {when $\mathcal{H}<0$}
and under the complete stability  {when $\mathcal{H}>0$}.
The concrete examples are shown in TABLE \ref{table_stable_sol}.

\subsubsection*{Bouncing solutions with asymptotic Milne spacetime} 
The solutions $\mathcal{B}^{[0;-1]}$ and $\mathcal{B}_{\mathrm{BC}}^{[0;-1]}$ correspond to this case. 
Recall that these types of singularity-free solutions cannot be stabilized in the projectable HL theory.
Since the Hubble parameter is dropped as $t^{-1} \sim a^{-1}$, 
the effect of the squared effective masses do not suppressed by the Hubble friction, 
that is, the values of $\mathcal{M}^2/H^2$ are negative constants and do not 
approach to zero at asymptotic Milne regime.

On the other hand, we can discover the stable solutions in the non-projectable HL theory 
because it is possible to keep the values of $\mathcal{M}^2_{(Q)}$ to be positive during the whole evolution.
In FIG. \ref{M2H2_L0}, a typical example of such a solution is shown.
\begin{figure}[tb!]
\begin{center}
\includegraphics[width=80mm]{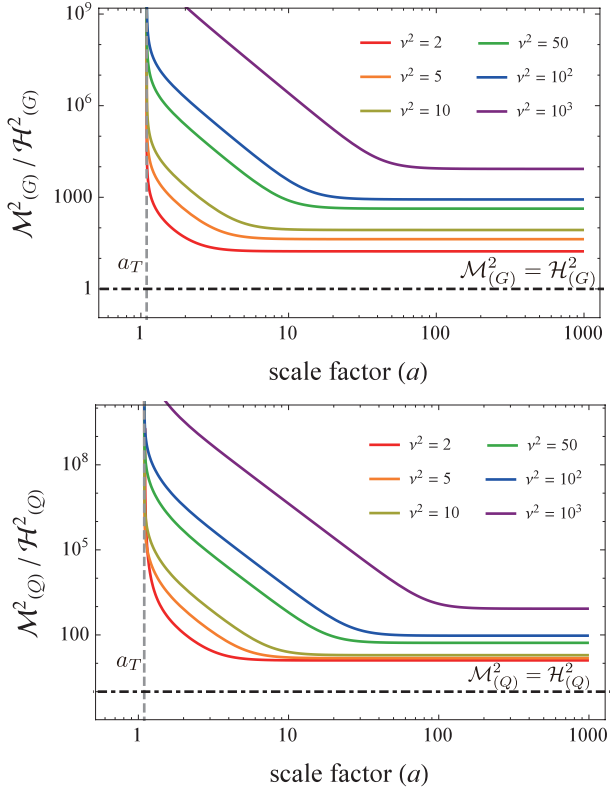} 
\end{center}
\caption{
The typical example of the ratios of the squared effective mass to
 {the squared effective friction coefficient $\mathcal{M}^2/\mathcal{H}^2$} 
in the bouncing solution with asymptotic Milne spacetime.
The top and the bottom figures stands for  {$\mathcal{M}_{(G)}^2/\mathcal{H}_{(G)}^2$ 
and $\mathcal{M}_{(Q)}^2/\mathcal{H}_{(Q)}^2$}, respectively.
The example solution is $\mathcal{B}^{[0;-1]}$ whose values of coupling constants are given in TABLE \ref{table_stable_sol} (ii).
The red, orange, yellow, green, blue and purple curves indicate the evolutions with 
$\nu^2=2, 5, 10, 50, 10^2$ and $10^3$, respectively ($\nu^2 =n^2+1$).
The horizontal dot-dashed black lines indicate {$\mathcal{M}^2=\mathcal{H}^2$}.
The vertical dashed gray line indicates the bouncing radius.
}
\label{M2H2_L0}
\end{figure}
The figure shows that $\mathcal{M}^2/\mathcal{H}^2$ of both tensor and scalar perturbations keep the positive values
and always greater than unity.
Thus, the  {$\mathcal{H}$-accelerated} instability does not occur for any initial value of the scale factor $a_{\mathrm{ini}}$.
As far as our numerical analysis is concerned, such behavior can also be seen for any possible perturbation mode $\nu$.
Therefore we regard our example as a solution under complete stability  {when $\mathcal{H}>0$} 
and under  {$\mathcal{M}^2$-dominated stability}  {when $\mathcal{H}<0$} for any $a_{\mathrm{ini}}$.

One may wonder whether the squared effective masses truly suppress 
the effects of  {$\mathcal{H}$-acceleration}
under the  {$\mathcal{M}^2$-dominated stability}, because we do not mention any explicit criterions
of the ratio $\mathcal{M}^2/\mathcal{H}^2$ so far.
Thus, we additionally discuss the dynamical evolutions of each perturbation modes
to investigate the growth of the perturbation amplitudes caused by the  {$\mathcal{H}$-acceleration}. 

To clarify the effects, we numerically solve the perturbation equations of motion
(\ref{pert_eom_st}). 
The initial conditions are given at $a_{\mathrm{ini}}=100 a_T$ so that
$\dot{h}_{(G)} =0 = \dot{h}_{(Q)} $. 
Since the equations of motion are linear with respect to $h_{(G)}$ and 
$h_{(Q)}$, 
the whole evolutions are proportional to the values of the initial conditions  $h_{(G) \mathrm{ini} }$ and $h_{(Q) \mathrm{ini} }$.
Therefore, we trace the ratios to each initial values, i.e., $h_{(G)} / h_{(G) \mathrm{ini} }$ and $h_{(Q)} / h_{(Q)\mathrm{ini} }$. 
In FIG \ref{perturbation_dynamics}, the time evolutions are shown.
\begin{figure}[tb]
\begin{center}
\includegraphics[width=80mm]{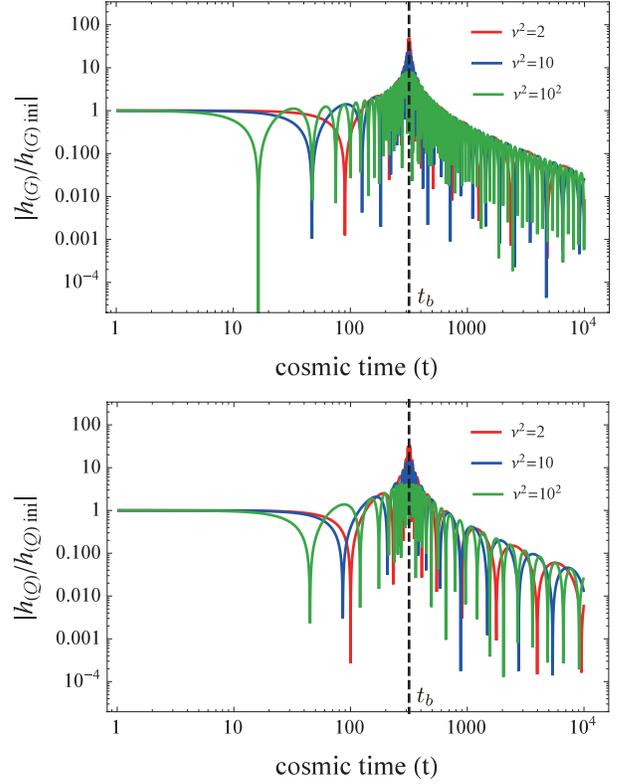} 
\end{center}
\caption{
The evolutions of perturbations in the bouncing solution $\mathcal{B}^{[0;-1]}$ where the coupling constants are
listed in TABLE~\ref{table_stable_sol} (ii).
In the top (bottom) figure, the evolution of tensor (scalar) perturbations are illustrated.
The initial conditions are selected by $a_{\mathrm{ini}}=100a_T\approx109.41$, and
$\dot{h}_{(G)}=0=\dot{h}_{(Q)}$.
The red, blue and green curves correspond to $\nu^2=2, 10$ and $100$, respectively.
The vertical dashed black line means the time of bouncing $t_b\approx317.19$.
}
\label{perturbation_dynamics}
\end{figure}
\begin{table}[h!]
\caption{
The maximum 
ratios $h_{(G)} / h_{(G) \mathrm{ini} }$ and $h_{(Q)} / h_{(Q) \mathrm{ini} }$ throughout the evolutions  for each perturbation modes
based on TABLE~\ref{table_stable_sol} (ii).
}
\begin{center}
\begin{tabular}{ccrrccc}
\hline \hline
\rule[-2mm]{0mm}{5mm}& $\nu^2$ &~$\max( h_{(G)} / h_{(G) \mathrm{ini} } )$ 
& $\max( h_{(Q)} / h_{(Q) \mathrm{ini} } )$  \\
\hline \hline
\rule[-2mm]{0mm}{5mm}&  $2$ & {  49.425} ~~~~~& {  32.907} ~~~~~\\ 
\rule[-2mm]{0mm}{5mm}&  $5$ & {  33.761} ~~~~~& {  20.835} ~~~~~\\ 
 \rule[-2mm]{0mm}{5mm}&   $10$  & {  24.876} ~~~~~& {  13.945} ~~~~~\\
 \rule[-2mm]{0mm}{5mm}&  $50$ & {  11.612} ~~~~~& {  5.933} ~~~~~\\  
\rule[-2mm]{0mm}{5mm}&  $100$ & {  8.263} ~~~~~& {  4.132} ~~~~~\\ 
\rule[-2mm]{0mm}{5mm}&  ${  300}$ & {  4.795} ~~~~~& {  2.308} ~~~~~\\ 
\hline \hline 
\end{tabular}
\label{enhancement_table}
\end{center}
\end{table} 
This result is based on the solution $\mathcal{B}^{[0; -1]}$ whose coupling constants
are given in TABLE~\ref{table_stable_sol} (ii), thus,
also corresponds with FIG \ref{M2H2_L0}.
From the figure, we find that the perturbation amplitudes are enhanced around the bouncing point
even if the solution is under the  $\mathcal{M}^2$-dominated stability.
Thus, we should clarify the growth rates of the perturbations.
In TABLE \ref{enhancement_table}, we show the detail data of the maximum 
ratios $h_{(G)} / h_{(G) \mathrm{ini} }$ and $h_{(Q)} / h_{(Q)\mathrm{ini} }$ 
throughout the evolutions  for each perturbation modes.
From the data, the amplitude possibly grows up to about $50$ times and 
tends to be higher as the perturbation mode $\nu^2$ is lower.
Since the large perturbation mode $\nu^2$ corresponds to
the heavy effective squared mass $\mathcal{M}^2$, 
the result seems appropriate.

\subsubsection*{Bouncing solutions with asymptotic de Sitter spacetime} 
The solutions $\mathcal{B}^{[1;1]}$, $\mathcal{B}^{[1;1]}_{\mathrm{BC}}$, $\mathcal{B}^{[1;1]}_{\mathrm{O}}$
with $a \geq a_T$, $\mathcal{B}^{[1;-1]}$ and $\mathcal{B}^{[1;-1]}_{\mathrm{BC}}$ correspond to this case. 
Actually, the previous work based on the projectable HL theory has shown that we cannot construct a solution 
under the complete stability. 
Instead of this, it is possible to {find solutions under the {$\mathcal{H}$}-suppressed stability
due to the negative $\mathcal{M}_{(Q)}^2$ in infrared regime.

On the other hand, it turns out that solutions under the complete stability can be realized in the non-projectable case.
As an example, we show the evolutions of the $\mathcal{M}^2/\mathcal{H}^2$ in FIG. \ref{M2H2_bounce_pL}.
\begin{figure}[bt!]
\begin{center}
\includegraphics[width=80mm]{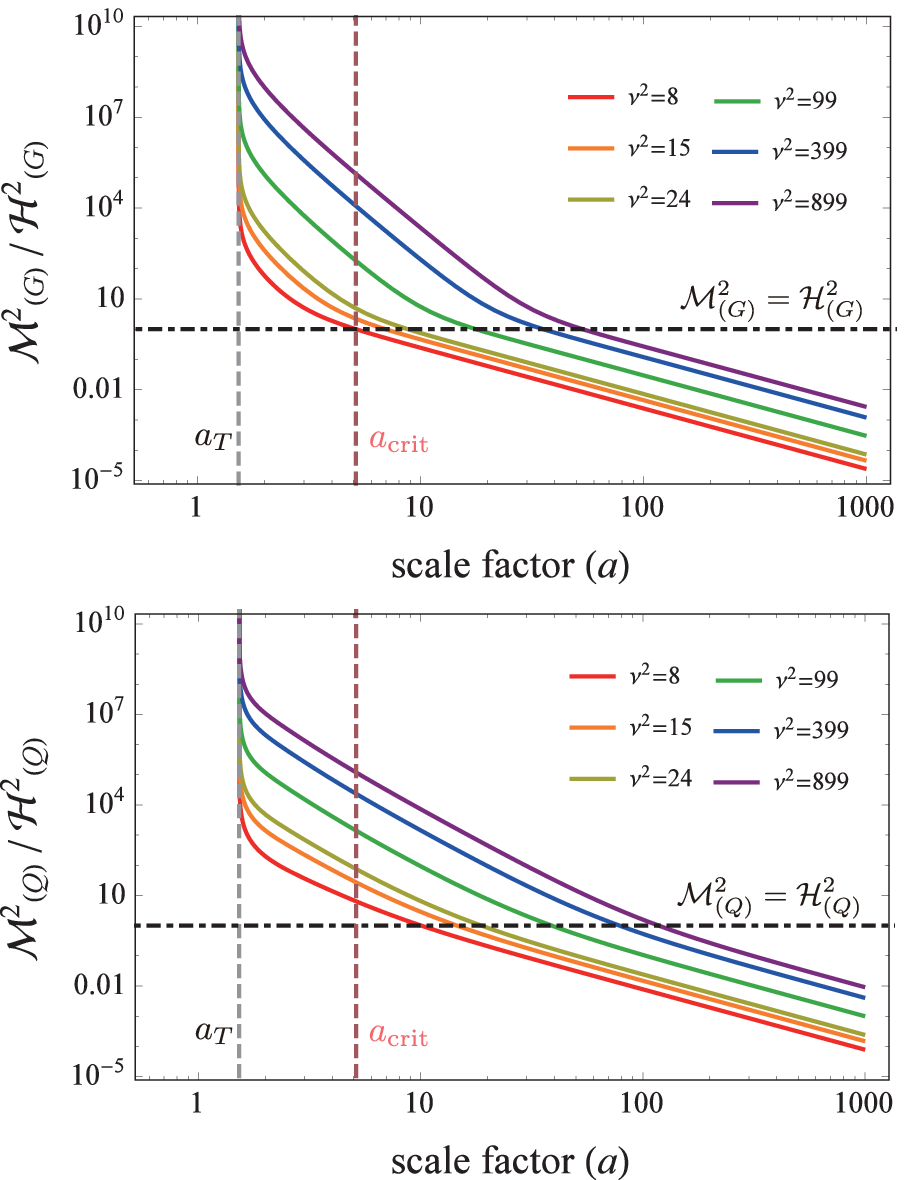} 
\end{center}
\caption{
The typical example of the ratios of the squared effective mass to 
the squared effective friction coefficient $\mathcal{M}^2/\mathcal{H}^2$
in the bouncing solution with asymptotic de Sitter spacetime.
The top and the bottom figures stands for  {$\mathcal{M}_{(G)}^2/\mathcal{H}_{(G)}^2$ and $\mathcal{M}_{(Q)}^2/\mathcal{H}_{(Q)}^2$}, respectively.
The example solution is $\mathcal{B}^{[1;1]}$ whose values of coupling constants are given 
in TABLE \ref{table_stable_sol} (iv).
The red, orange, yellow, green, blue and purple curves indicate the evolutions with 
$\nu^2=8, 15, 24, 99, 399$ and $899$, respectively ($\nu^2 =n^2-1$).
The horizontal dot-dashed black lines indicate {$\mathcal{M}^2=\mathcal{H}^2$}.
The vertical dashed gray line gives the bouncing radius,
while the vertical dashed pink line shows the critical scale factor {$a_{\mathrm{crit}}=5.118$}.
}
\label{M2H2_bounce_pL}
\end{figure}
As can be seen from the figure, both tensor and scalar perturbations keep the positive values during whole evolution
of the solution. 
As far as our analysis is concerned, the squared effective masses of tensor and scalar perturbations are monotonically 
increasing functions with respect to the perturbation mode $\nu^2$.
Therefore we conclude this solution is under complete stability for any possible $\nu^2$.

The magnitude relationships between $\mathcal{M}^2$ and $\mathcal{H}^2$ are also analyzed.
As we discussed, both $\mathcal{M}^2_{(G)}$ and $\mathcal{M}^2_{(Q)}$ drop as $a^{-2}$, 
in contrast, the  {effective friction coefficients are} settled to a constant.
Thus, there must exist a certain value of the scale factor such that any one of $\mathcal{M}^2$ 
is equal to $\mathcal{H}^2$.
We define such a value of scale factor as a {\it critical scale factor} $a_{\mathrm{crit}}$.
In our numerical example FIG. \ref{M2H2_bounce_pL}, the critical scale factor is $a_{\mathrm{crit}}=5.118$
at which the lowest $\nu^2$ mode of the tensor perturbation shows
 $\mathcal{M}_{(G)}^2=\mathcal{H}^2_{(G)}$
and $\mathcal{M}_{(G)}^2$ is weaker than $\mathcal{H}_{(G)}^2$ for $a>a_{\mathrm{crit}}$.
Thus, in order to prevent $\mathcal{H}$-accelerated instability, the initial value of the scale factor $a_{\mathrm{ini}}$ needs to be 
smaller than that critical value.

We also mention the growth of the perturbations around the bouncing point.
Actually, it is found that the behavior is not quite different from the previous case with asymptotic Milne spacetime.
It is reasonable because the effect of a cosmological constant is expected to be relatively weaker than those of
higher order curvature terms.

\subsubsection*{Oscillating solutions} 
We examine the stabilities of the oscillating solutions, specifically, $\mathcal{O}^{[0;1]}$, $\mathcal{B}^{[1;1]}_{\mathrm{O}}$ for 
$a_{\mathrm{min}} \leq a \leq a_{\mathrm{max}}$, 
$\mathcal{O}^{[-1;1]}$, $\mathcal{O}^{[-1;-1]}$ and $\mathcal{O}^{[-1;-1]}_{\mathrm{BC}}$ for 
$a_{\mathrm{min}} \leq a \leq a_{\mathrm{max}}$. 
In this case, we also construct solutions whose squared effective masses are positive and 
dominate the {$\mathcal{H}$-term}  {within} the possible ranges of the scale factor. 
A typical example is shown in FIG. \ref{M2H2_oscillating_LN}.
\begin{figure}[bt!]
\begin{center}
\includegraphics[width=80mm]{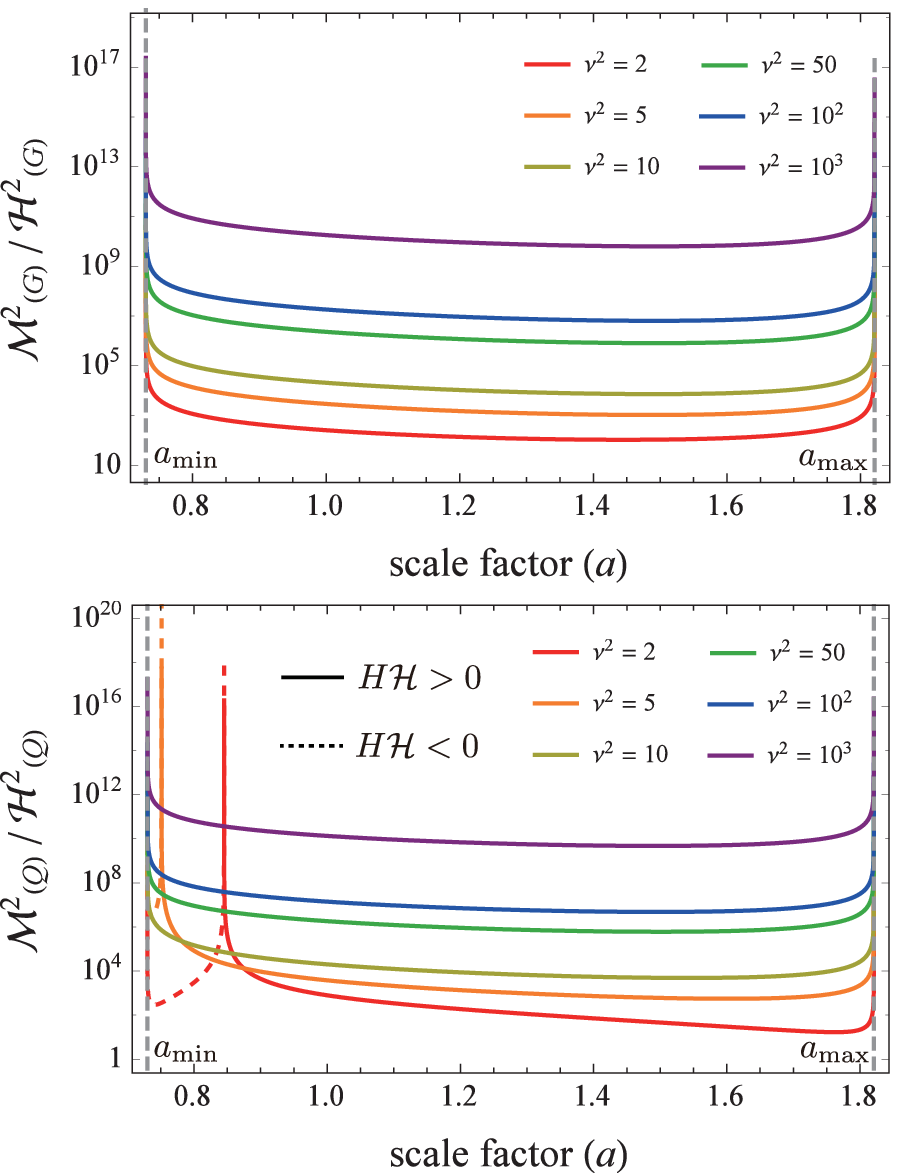} 
\end{center}
\caption{
 {
The typical example of the ratios of the squared effective mass to 
 {the squared effective friction coefficient $\mathcal{M}^2/\mathcal{H}^2$} 
in the oscillating solution.
The top and the bottom figures stands for 
 {$\mathcal{M}_{(G)}^2/\mathcal{H}_{(G)}^2$ and $\mathcal{M}_{(Q)}^2/\mathcal{H}_{(Q)}^2$}, respectively.
The example solution is $\mathcal{O}^{[-1;-1]}$ whose values of coupling constants are given in TABLE \ref{table_stable_sol} (x).}
The red, orange, yellow, green, blue and purple curves indicate the evolutions with 
$\nu^2=2, 5, 10, 50, 10^2$ and $10^3$, respectively ($\nu^2 =n^2+1$).
 {The solid and dashed colored curves represent the evolutions with $H\mathcal{H}>0$ and $H\mathcal{H}<0$, respectively.}
The vertical dashed gray lines indicate the maximum and minimum radii of the oscillation.
}
\label{M2H2_oscillating_LN}
\end{figure}
In our analysis, $\mathcal{M}^2$ of all perturbation modes are positive and greater than {$\mathcal{H}^2$}.
It means that the solution is under the $\mathcal{M}^2$-dominated stability  {when $\mathcal{H}<0$}
and under the complete stability  {when $\mathcal{H}>0$}.
It is remarkable that, in that case, the scalar perturbations with low $\nu^2$ show $H\mathcal{H}_{(Q)} <0$ 
around the bouncing point, which means that the scalar perturbations feel $\mathcal{H}$-friction slightly before the bounce and
receive $\mathcal{H}$-acceleration slightly after the bounce.

Additionally, we show the dynamics of the tensor and scalar perturbations 
in the oscillating solution in FIG.~\ref{perturbation_dynamics_osc}.
\begin{figure}[tb!]
\begin{center}
\includegraphics[width=80mm]{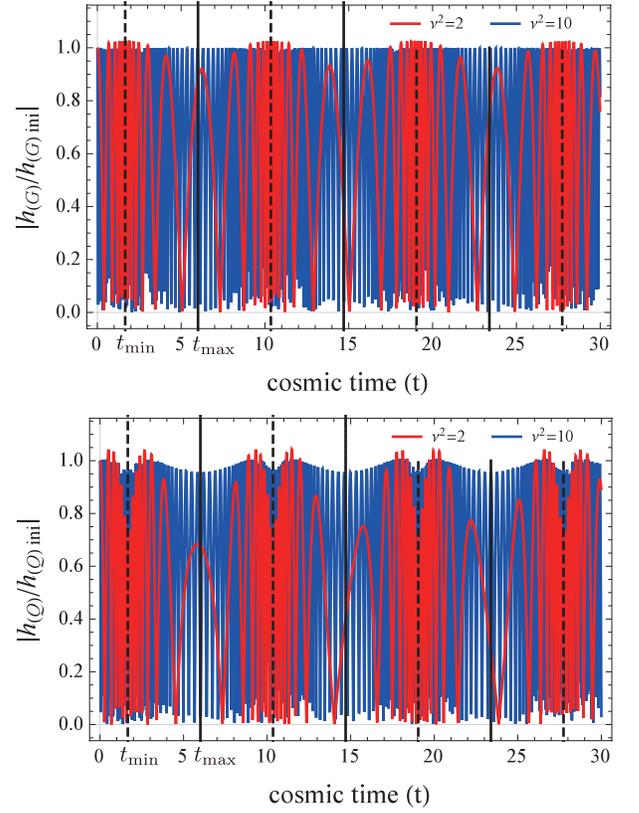} 
\end{center}
\caption{
The evolutions of perturbations in the oscillating solution $\mathcal{O}^{[-1;-1]}$ 
whose coupling constants are listed in TABLE~\ref{table_stable_sol} (x).
In the top (bottom) figure, the evolution of tensor (scalar) perturbations are illustrated.
The initial conditions are set to $a_{\mathrm{ini}}=(a_{\mathrm{min}}+a_{\mathrm{max}})/2\approx1.275$, and
$\dot{h}_{(G)}=0=\dot{h}_{(Q)}$.
The red and blue curves correspond to $\nu^2=2$ and $10$, respectively.
The dashed and solid black lines indicate the bouncing time and the recollapsing time, respectively.
The first bouncing and recollapsing times are given by $t_{\mathrm{min}}\approx1.660$ and $t_{\mathrm{max}}\approx6.006$.
The oscillating period of the scale factor is  $T \approx 8.692$.} 
\label{perturbation_dynamics_osc}
\end{figure}
From these figures, the typical amplitudes of the perturbations are almost constant 
even if the universe experiences a number of oscillations.
Thus, we conclude the solution is stable with respect to the linear perturbations.

\section{summary and discussions}
\label{conclusion}
We have investigated the stability of the singularity-free cosmological solutions based on the non-projectable HL theory
whose action is given by (\ref{HL_action}).
Since our aim is to remedy the infrared behaviors of the singularity-free solutions in the projectable HL theory,
we introduced only the single additional term $\Phi^2$ which is expected to be dominant in infrared limit.
It is remarkable that our gravitational action realizes the identical background solutions in FLRW spacetime
based on the projectable HL theory with $\mathcal{C}=0$.
Therefore, the bouncing solutions and the oscillating solutions which are induced by the higher order 
spatial derivative terms in the action
are also found as in the projectable case.

By considering the quadratic action, we discuss the stability of the singularity-free
solutions with respect to tensor and scalar perturbations.
The stabilities can be estimated by the sign of the coefficients $\mathcal{F}$ and $\mathcal{G}$ showed in (\ref{action_tensor}) and (\ref{action_scalar}). 
$\mathcal{F}<0$ corresponds to the ghost instability which is equivalent to a lack of the lowest energy state.
The case with $\mathcal{G}<0$ is known as the gradient instability which is interpreted as
a unstable behavior due to the negative squared effective mass.
However, it is possible to consider the case that the unstable behavior caused by $\mathcal{G}<0$ is suppressed by the effect of background dynamics.
As we showed in (\ref{pert_eom_st}), the stability of the perturbations can be
judged by the sign of the effective friction coefficients $\mathcal{H}$ and 
the effective squared masses $\mathcal{M}^2$, additionally 
the magnitude relationships between $\mathcal{H}^2$ and $|\mathcal{M}^2|$.
Thus, we introduced five types of stabilities and instabilities 
(i) negative-$\mathcal{M}^2$ instability, (ii) $\mathcal{H}$-accelerated instability, (iii) $\mathcal{M}^2$-dominated stability, 
(iv) $\mathcal{H}$-suppressed stability, and (v) complete stability.

The novel feature of the singularity-free cosmological solutions in the non-projectable HL theory
is that we can find the bouncing solutions which satisfy the 
$\mathcal{M}^2$-dominated stability condition {when $\mathcal{H}<0$}
and the complete stability condition {when $\mathcal{H}>0$}.
Such solutions cannot be constructed in the projectable HL theory,
that is, the squared effective masses must be negative in infrared region. 

We additionally investigate the stability of the oscillating solutions.
The solution (x) we have shown in Sec. \ref{section_numerical} is 
the stable solution which satisfy the $\mathcal{M}^2$-dominated stability condition {when $\mathcal{H}<0$}
and the complete stability condition {when $\mathcal{H}>0$}.
In fact, the typical amplitudes of the perturbations stay almost constant.

However, we would like to indicate that it is not impossible to construct 
an oscillating solution whose typical perturbation amplitude is exponentially enhanced 
even if either $\mathcal{M}^2$-dominated stability condition or 
the complete stability condition is satisfied.
That instability is caused by a {\it resonance}.
Whether the resonance is induced or not can be investigated 
by the Hill's method\cite{ref_Hill_eq} which is summarized in Appendix \ref{resonance}.
\begin{table*}[!t]
\caption{
An example of the coupling constants for the oscillating solution with resonance.
}
{
\begin{center}
\begin{tabular}{cc||c|c|c|c|c|c|c|c|c|c||c|c||c|ccc}
\hline \hline
\rule[-1mm]{0mm}{4mm}&~type~~&~$\Lambda$~~&~$\lambda$~~&~$g_2$~&~$g_3$~~&~$g_4$~&~$g_5$~&~$g_6$~&~~$g_7$~&
~~$g_8$~&~~$\varsigma$~&~~$g_r$~&~~$g_s$~&~~$a_{\mathrm{min}}$~&~~$a_{\mathrm{max}}$~\\
\hline \hline 
\rule[-3.5mm]{0mm}{9mm}~~&$\mathcal{O}^{[-1;-1]}$~&
~~$-1$~~&~~$3$~~&~~$\displaystyle{{1 \over 4}}$~~&~~$\displaystyle{-{1 \over 2}}$~~&~~$0$~~&~~$\displaystyle{{1 \over 20}}$~~&
~~$\displaystyle{-{1 \over 15}}$~~&~~$\displaystyle{-{1 \over 5}}$~~&~~$\displaystyle{{1 \over 5}}$~~&~~$\displaystyle{{3 \over 2}}$~~&~~$\displaystyle{{3 \over 2}}$~~&~~$1$~~
&~~0.629~~&~~1.832~~
 \\ 
\hline \hline 
\end{tabular}
\label{table_resonance}
\end{center}}
\end{table*}
We numerically show a example of the oscillating solution with resonance 
instability in FIG. \ref{hg_dyn}
and the coupling constants of the solution is shown in TABLE~\ref{table_resonance}.
Note that this solution is satisfied both $\mathcal{M}^2$-stability condition 
and complete stability condition (see FIG.~\ref{M2H2_resonance}).
In that case, certain modes of the tensor perturbation show unstable behavior.
The Hill's method indicates that the degree of enhancement is characterized by $\epsilon_\pm$ defined in (\ref{def_epsilon}).
If $|\epsilon_{\pm}|$ exceeds 1, the amplitude of the corresponding perturbation mode exponentially grows.
In fact, the solution includes certain tensor perturbation modes with $|\epsilon_{\pm}|>1$ 
as we show in FIG.\ref{ep_nu_in}.
It shows that even if all stability conditions we have discussed are satisfied, 
the oscillating solution is possibly unstable due to the resonance.

 \begin{figure}[!b]
\begin{center}
\includegraphics[width=80mm]{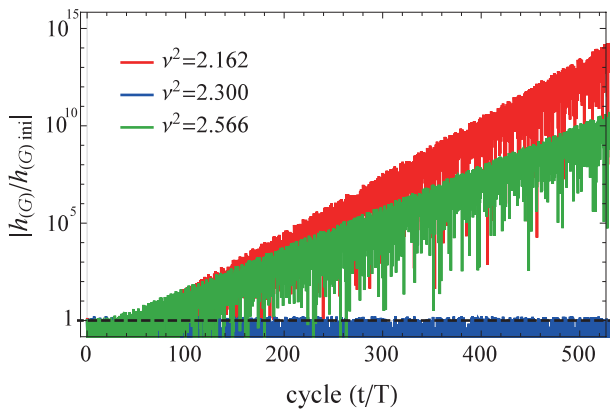}
\end{center}
\caption{
The evolutions of the tensor perturbations in the oscillating solution 
where the coupling constants are listed in TABLE~\ref{table_resonance}.
The oscillating period of the scale factor is  $T \approx  8.723$.
The $\nu^2=2.162$ (red) and $2.556$ (green) modes show the resonance, while the $\nu=2.300$ (blue) mode is stable.
}
\label{hg_dyn}
\end{figure}

The oscillating solutions with resonance instability can be discovered 
if we set the coupling constant $g_8$ to be positive, however, relatively smaller than
that of the stable solutions listed in TABLE \ref{table_stable_sol}
as far as our numerical analysis is concerned.
One may notice that such a manipulation corresponds to consider the small effective squared mass.
See (\ref{Gg}) and (\ref{G_Q}), 
it is found that ultraviolet dominant terms $g_8 \nu^6/a^6$ decrease.
Thus, it is natural to consider that the heavy effective squared masses prevent 
the oscillating solution from the resonance instability. 
Actually, we can see from FIG.\ref{ep_nu_in} that the degree of enhancement 
$|\epsilon_\pm|$ approaches unity as the perturbation mode $\nu^2$ increases. 
We would like to stress that the resonance instability is mainly the problem in open FLRW spacetime.
Since the perturbation mode $\nu^2$ takes discrete number greater than or equal to eight in closed FLRW spacetime, 
the resonance instability is not quite problematic.
In general terms, the positive heavy squared effective masses 
$\mathcal{M}^2$ is preferred to avoid the instabilities due to the background dynamics, 
i.e., $\mathcal{H}$-accelerated instability and the resonance instability.
Thus, the higher order curvature terms with $z=3$ are significant to stabilize the spacetime 
as well as inducing the bounce.

 \begin{figure}[!b]
\begin{center}
\includegraphics[width=80mm]{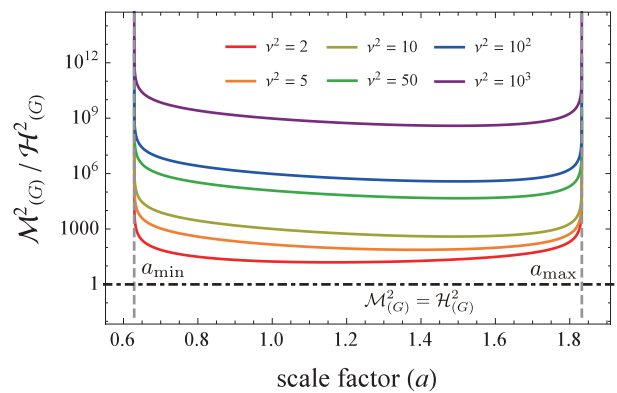}
\end{center}
\caption{
The typical example of the ratios of the squared effective mass to 
 the squared effective friction coefficient $\mathcal{M}_{(G)}^2/\mathcal{H}_{(G)}^2$
 in the oscillating solution with resonance instability.
The example solution is $\mathcal{O}^{[-1;-1]}$ whose values of coupling constants are given in TABLE \ref{table_resonance}.
The red, orange, yellow, green, blue and purple curves indicate the evolutions with 
$\nu^2=2, 5, 10, 50, 10^2$ and $10^3$, respectively ($\nu^2 =n^2+1$).
The horizontal dot-dashed black lines indicate {$\mathcal{M}_{(G)}^2=\mathcal{H}_{(G)}^2$}.
The vertical dashed gray lines indicate the maximum and minimum radii of the oscillation.
}
\label{M2H2_resonance}
\end{figure}

 \begin{figure}[!bht]
\begin{center}
\includegraphics[width=80mm]{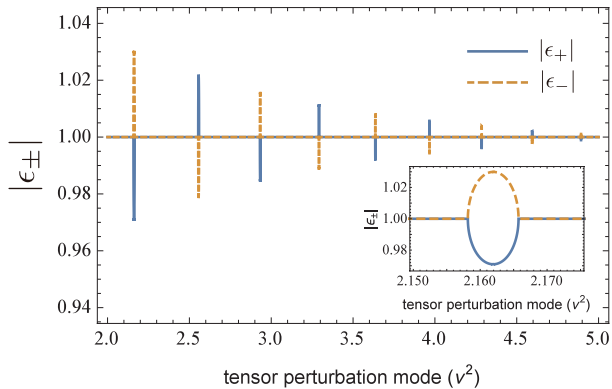}
\end{center}
\caption{
The $\epsilon_\pm$ dependence on the tensor perturbation mode $\nu^2$.
The coupling constants are listed in TABLE \ref{table_resonance}.
The solid blue line shows $|\epsilon_+|$, while the dashed orange line gives $|\epsilon_-|$.
$|\epsilon_-|$ reaches a maximum($\approx 1.030$) at $\nu^2\approx2.162$
 and $|\epsilon_+|$ reaches a maximum($\approx 1.021$) at $\nu^2\approx2.556$.
This system causes the resonance around those modes.
}
\label{ep_nu_in}
\end{figure}

We also mention the stability of singularity-free solutions 
based on the {\it general covariant} HL theory
which is an alternative modification  to remedy the behavior of scalar perturbation\cite{U1_PHL1}.
In this theory, the scalar propagating degree of freedom is eliminated by 
the additional fields, 
whereas the tensor perturbations are not modified.
It is notable that the identical background solutions in FLRW spacetime to 
those of projectable and non-projectable HL theory can be reproduced 
under certain conditions\cite{U1_PHL2,U1_NPHL1,U1_NPHL2}.
Therefore the stability analyses we have performed are also valid 
in the general covariant HL theory.
For example, the background solutions with $g_3 =0$ hold positive squared effective
masses if the conditions we have explicitly shown in Sec. \ref{tensor_stability} are satisfied.
Then, we can construct the stable singularity-free solutions 
only by examining $\mathcal{M}^2_{\mathrm{(G)}}/H^2$.

The perturbative approach we have performed is based on a postulate.
The effect of the backreaction can be ignored.
In other words,  the background dynamics are never affected by the perturbations.
Of course, that assumption is not trivial.
If the perturbation amplitudes are much enhanced, it is possible that the perturbation fields 
impinge on the background dynamics.
Our numerical analysis shows that the typical amplitudes of the perturbations are enhanced around bouncing point.
The degree of enhancement tends to be larger if $| \mathcal{M}^2 |$ is small.
Therefore one may wonder how much the perturbation amplitude is allowed not to affect to the background dynamics.
For certain perturbation mode, we can give an estimation. 
In fact, the tensor perturbations with $\nu^2 =8$ in closed FLRW spacetime include the homogeneous modes, 
which corresponds the Bianchi-type IX spacetime whose anisotropy is small (in detail,  see \cite{previous_stability_bounce}).
Thus it might be possible to estimate the effect of backreaction by considering more general background 
including inhomogeneity and anisotropy.

\section*{Acknowledgments}

The authors would like to thank K.~Maeda and K.~Aoki for 
their insightful comments and discussions.
MF and SS are grateful to the Early Bird Program from Waseda Research Institute for Science and Engineering, 
Grant-in-Aid for Young Scientists.
The work of S. M. was supported in a part by Grants-in-Aid from the Scientific Research Fund of the Japan Society 
for the Promotion of Science (No.18J11983).


\appendix

\section{resonance in the oscillating solution}
\label{resonance}

In the oscillating solutions, the perturbations feel periodical external forces, 
which may causes a resonance and the rapid growth of the perturbations.
To clarify whether the resonance is induced or not, we examine the perturbation equations by 
adopting Hill's method.
Since the procedures are completely the same between tensor and scalar perturbations,
the subscripts which represent tensor and scalar modes are abbreviated in what follows.

To rewrite the equations into the Hill's form, we consider a field redefinition as
\begin{eqnarray}
\hat{h} := \sqrt{ a^3 \mathcal{F}}\, h.
\end{eqnarray}
Then,  the perturbation equations are transformed as 
\begin{eqnarray}
\frac{d^2\hat{h}}{dt^2}&+\omega^2(t)\hat{h}=0 \,, \label{Hilleq}
\end{eqnarray}
with
\begin{eqnarray}
\omega^2(t):=\frac{1}{4}\left[\frac{\dot{\mathcal{F}}^2}{\mathcal{F}^2}-\frac{2(3H\dot{\mathcal{F}}+\ddot{\mathcal{F}})}{\mathcal{F}}
-9H^2 -6\dot{H} + 4\mathcal{M}^2  \right] \,, 
\notag \\
\end{eqnarray}
Since the effective squared angular frequency $\omega^2(t)$ is essentially a function which depend only on the scale factor, 
the oscillating period is expected to be the same as that of the scale factor.
We further introduce the following equation in matrix expression as
\begin{eqnarray}
\frac{d}{dt}X(t)
 &=&
 \left(
    \begin{array}{cc}
      0 & 1 \\
      -\omega^2(t)&0 
    \end{array}
  \right)
  X(t)\,,~\\ \notag \\
X(t) &:=& 
\begin{pmatrix}
\hat{h}(t)   \\
  \dot{\hat{h}}(t)
 \end{pmatrix}\,,
\end{eqnarray}
Here  $\hat{h}(t)$ is an arbitrary perturbation function and $X(t)$ is a real solution vector for the Hill equation (\ref{Hilleq}).

Let $T$ be a period which is characterized by the effective squared angular momentum $\omega^2$, 
then, the evolution of $X$ is expressed as 
\begin{align}
X(t+T)=\mathcal{A} X(t) \,, \label{def_A}
\end{align}
where, $\mathcal{A}$ is a $2 \times 2$ matrix which represents the time translation $t \to t + T$.
Suppose \{$\epsilon_+, X_{\epsilon_+}$\} and \{$\epsilon_-, X_{\epsilon_-}$\}  are two independent 
eigensystems of matrix $\mathcal{A}$, 
where, $\epsilon_\pm$ are eigenvalues  and $X_{\epsilon_\pm}$ are corresponding eigenvectors.
Then, (\ref{def_A}) can be rewritten as some linear combination of the following relations: 
\begin{eqnarray}
 X_{\epsilon_\pm}  (t+T) = \epsilon_\pm X_{\epsilon_\pm} (t) \,.
 \label{def_epsilon}
 \end{eqnarray}
 Therefore, we find that both $|\epsilon_+| \leq 1$ and $|\epsilon_-| \leq 1$ are required 
 to suppress the resonance of the perturbations.

 To specify the explicit conditions for $|\epsilon_\pm| \leq 1$, 
 we introduce two independent complex vector $X_1$ and $X_2$
 whose initial conditions are
\begin{align}
X_1(0)&= 
  \left(
    \begin{array}{c}
      \hat{h}_1(0) \\
      \dot{\hat{h}}_1(0)
    \end{array}
  \right)
  = \left(
    \begin{array}{c}
      1 \\
      0 
    \end{array}
  \right)\,,~ \label{X1} \\
X_2(0)&=
   \left(
    \begin{array}{c}
      \hat{h}_2(0) \\
      \dot{\hat{h}}_2(0)
    \end{array}
  \right)
=  \left(
    \begin{array}{c}
      0 \\
      1 
    \end{array}
  \right).   \label{X2}
\end{align}
Any solution vector can be constructed by the linear combination of $X_1$ and $X_2$.
Substituting (\ref{X1}) and (\ref{X2}) into (\ref{def_A}), we find
\begin{align}
\mathcal{A}=\left(
    \begin{array}{cc}
       \hat{h}_1(T)  &   \hat{h}_2(T)  \\
       \dot{\hat{h}}_1(T)  & \dot{\hat{h}}_2(T)  
    \end{array}
  \right).
\end{align}
 Solving the Hill equation (\rm{\ref{Hilleq}}) with these initial conditions, $\mathcal{A}$ can be estimated.
The eigenvalues of $\mathcal{A}$ are obtained from the following equation.
\begin{align}
\epsilon^2 -(\mathrm{tr} \mathcal{A}) \epsilon +\mathrm{det} \mathcal{A}=0.
\end{align}
Since the Hill equation (\rm{\ref{Hilleq}}) and the initial conditions (\ref{X1}) and (\ref{X2}) give $\mathrm{det}\mathcal{A}=1$,
the eigenvalues $\epsilon_\pm$ are given by
\begin{align}
\epsilon_\pm=\frac{ \mathrm{tr} \mathcal{A}\pm\sqrt{( \mathrm{tr} \mathcal{A})^2-4}}{2}.
\end{align}
One can see that $| \mathrm{tr} \mathcal{A} | \leq 2$ is equivalent to $|\epsilon_\pm| = 1$.
When $\epsilon_\pm$ is a complex number, the perturbations correspond to the real part of $X(t)$.
For the case with $| \mathrm{tr} \mathcal{A} | > 2$, any one of $|\epsilon_\pm|$ is greater than unity.
Thus, we conclude that if $| \mathrm{tr} \mathcal{A} | \leq 2$, the perturbation keeps oscillating forever without growing or decaying.
On the other hand,  the perturbation shows an exponential instability if $| \mathrm{tr} \mathcal{A} | > 2$.



\begin{thebibliography}{99}

 \bibitem{singularity_theorem}
 R.~Penrose, Phys. Rev. Lett. {\bf 14}, 57 (1965); 
 S.~W.~Hawking, Proc. R. Soc., A {\bf 300}, 187 (1967); 
 S.~W.~Hawking and R.~Penrose, Proc. R. Soc., A {\bf 314}, 529 (1970); 
 S.~W.~Hawking and G.~F.~R.~Ellis, {\it The Large Scale Structure of Space-Time} (Cambridge Univ., Cambridge, 1973).

\bibitem{inflation_not_complete}
A.~Borde and A.~Vilenkin,
 ``Eternal inflation and the initial singularity",
Phys. Rev. Lett. {\bf 72}, 3305 (1994);
A.~Borde, A.~H.~Guth and A.~Vilenkin,
 ``Inflationary Spacetimes Are Incomplete in Past Directions",
Phys. Rev. Lett. {\bf 90}, 151301 (2003) [arXiv:0110012[gr-qc]].

\bibitem{braneworld_bounce}
J.~Khoury, B.~A.~Ovrut, P.~J.~Steinhardt and N.~Turok,
``Ekpyrotic universe: Colliding branes and the origin of the hot big bang", 
Phys. Rev. D {\bf 64}, 123522 (2001) [arXiv:0103239[hep-th]];
J.~Khoury, B.~A.~Ovrut, N.~Seiberg, P.~J.~Steinhardt and N.~Turok,
``From big crunch to big bang", 
Phys. Rev. D {\bf 65}, 086007 (2002) [arXiv:0108187[hep-th]].

\bibitem{string_gas}
R.~Brandenberger and C.~Vafa,
``Superstrings in the early universe", 
Nucl. Phys. B {\bf 316}, 391 (1989);
A.~Nayeri, R.~Brandenberger and C.~Vafa,
``Producing a Scale-Invariant Spectrum of Perturbations in a Hagedorn Phase of String Cosmology", 
Phys. Rev. Lett. {\bf 97}, 021302 (2006) [arXiv:0511140[hep-th]];
R.~Brandenberger, R.~Costa, G.~Franzmann and A.~Weltman,
``Point particle motion in double field theory and a singularity-free cosmological solution", 
Phys. Rev. D {\bf 97}, 063530 (2018) [arXiv:1710.02412[hep-th]].

\bibitem{LQG_bounce1}
D.~J.~Mulryne, R.~Tavakol, J.~E.~Lidsey and G.~F.~R.~Ellis,
``An emergent universe from a loop", 
Phys. Rev. D {\bf 71}, 123512 (2005) [arXiv:0502589[astro-ph]].
\bibitem{LQG_bounce2}
I.~Agullo, A.~Ashtekar and W.~Nelson,
``The pre-inflationary dynamics of loop quantum cosmology: confronting quantum gravity with observations", 
Class. Quant. Grav. {\bf 30}, 085014 (2013) [arXiv:1302.0254[gr-qc]].
\bibitem{LQG_bounce3}
E.~W.~Ewing,
``Ekpyrotic loop quantum cosmology", 
JCAP {\bf 1308}, 015 (2013) [arXiv:1306.6582[gr-qc]].


\bibitem{genesis1}
P.~Creminelli, A.~Nicolis and E.~Trincherini,
``Galilean genesis: an alternative to inflation", 
JCAP {\bf 1011}, 021 (2010) [arXiv:1007.0027[hep-th]].
\bibitem{genesis2}
D.~A.~Easson, I.~Sawicki, and A.~Vikman,
``G-Bounce", 
JCAP {\bf 1111}, 021 (2011) [arXiv:1109.1047[hep-th]].
\bibitem{genesis3}
S.~Nishi and T.~Kobayashi, 
``Generalized galilean genesis", 
JCAP {\bf 1503}, 057 (2015) [arXiv:1501.02553[hep-th]].
\bibitem{genesis4}
T.~Kobayashi, M.~Yamaguchi and J.~Yokoyama,
``Galilean creation of the inflationary universe", 
JCAP {\bf 1507}, 017 (2015) [arXiv:1504.05710[hep-th]].

\bibitem{Galileon_bounce}
A.~Ijjas and P.~J.~Steinhardt,
``Classically Stable Nonsingular Cosmological Bounces", 
Phys. Rev. Lett. {\bf 117}, 121304 (2016) [arXiv:1606.08880[gr-qc]];
A.~Ijjas and P.~J.~Steinhardt,
``Fully stable cosmological solutions with a non-singular classical bounce", 
Phys. Lett. B {\bf 764}, 289 (2017) [arXiv:1609.01253[gr-qc]].

\bibitem{Galileon_bounce2}
A.~Ijjas,
``Space-time slicing in Horndeski theories and its implications for non-singular bouncing solutions",
JCAP {\bf 1802}, 007 (2018) [arXiv:1710.05990[gr-qc]].

\bibitem{Galileon_bounce3}
D.~A.~Dobre, A.~V.~Frolov, J.~T.~G.~Ghersi, S.~Ramazanov, and A.~Vikman,
``Unbraiding the Bounce: Superluminality around the Corner",
JCAP {\bf 1803}, 020 (2018) [arXiv:1712.10272[gr-qc]].

\bibitem{non-local_bounce}
 T.~Biswas, E.~Gerwick, T.~Koivisto and A.~Mazumdar, 
 ``Towards Singularity and Ghost-free Theories of Gravity"
 Phys. Rev. Lett. {\bf 108}, 031101 (2012) [arXiv:1110.5249[gr-qc]];
 T. Biswas, A.~S.~Koshelev, A.~Mazumdar and S.~Y.~Vernov,
``Stable bounce and inflation in non-local higher derivative cosmology"
JCAP {\bf 08}, 024 (2012) [arXiv:1206.6374[astro-ph.CO]];
\'{A}. de la Cruz-Dombriz, F. J. M. Torralba, and A. Mazumdar,
``Non-singular and ghost-free infinite derivative gravity with torsion"
arXiv:1812.04037[gr-qc].


\bibitem{no-go_Horndeski}
T.~Kobayashi,
``Generic instabilities of non-singular cosmologies in Horndeski theory: a no-go theorem",
Phys. Rev. D {\bf 94}, 043511 (2016) [arXiv:1606.05831[hep-th]].

\bibitem{no-go_multi}
S.~Akama and T.~Kobayashi,
``Generalized multi-Galileons, covariantized new terms, and the no-go theorem for non-singular cosmologies"
Phys. Rev. D {\bf 95}, 064011 (2017) [arXiv:1701.02926[hep-th]]

\bibitem{no-go_non-flat}
S.~Akama and T.~Kobayashi,
``General theory of cosmological perturbations in open and closed universes from the Horndeski action",
arXiv:1810.01863[gr-qc].

\bibitem{no-go_papers}
R.~Kolevatov, and S.~Mironov,
``Cosmological bounces and Lorentzian wormholes in Galileon theories with an extra scalar field",
Phys. Rev. D {\bf 94}, 123516 (2016) [arXiv:1607.04099[hep-th]].

\bibitem{Horndeski_theory}
G.~W.~Horndeski, 
``Second-order scalar-tensor field equations in a four-dimensional space", 
Int. J. Theor. Phys. {\bf 10}, 363 (1974); 
T.~Kobayashi, M.~Yamaguchi and J.~Yokoyama, 
``Generalized G-inflation: Inflation with the most general second-order field equations", 
Prog. Theor. Phys. {\bf 126}, 511 (2011) [arXiv:1105.5723 [hep-th]].

\bibitem{bounce_EFT}
Y.~Cai, Y.~Wan, H-G.~Li, T.~Qiu and Y-S.~Piao, 
``The Effective Field Theory of nonsingular cosmology", 
JHEP {\bf 1701}, 090 (2017) [arXiv:1610.03400[gr-qc]];
Y.~Cai, Y.~Wan, H-G.~Li, T.~Qiu and Y-S.~Piao, 
``The Effective Field Theory of nonsingular cosmology:II", 
Eur. Phys. J. C {\bf 77}, 369 (2017) [arXiv:1701.04330[gr-qc]].


\bibitem{no-go_loopholes1}
P.~Creminelli, D.~Pirtskhalava, L.~Santoni, and E.~Trincherini,
``Stability of Geodesically Complete Cosmologies",
JCAP {\bf 1611}, 047 (2016) [arXiv:1610.04207[hep-th]].

\bibitem{no-go_loopholes2}
S.~Banerjee, and E.~N.~Saridakis,
``Bounce and cyclic cosmology in weakly broken galileon theories",
Phys. Rev. D {\bf 95}, 063523 (2017) [arXiv:1604.06932[gr-qc]].

\bibitem{no-go_loopholes3}
R.~Kolevatov, S.~Mironov, N.~Sukhov, and V.~Volkova, 
``Cosmological bounce and Genesis beyond Horndeski",
JCAP {\bf 1708}, 038 (2017) [arXiv:1705.06626[hep-th]].

\bibitem{Horava}
P.~Ho\v{r}ava, 
``Quantum gravity at a Lifshitz point", 
Phys. Rev. D {\bf 79}, 084008 (2009) [arXiv:0901.3775[hep-th]].


\bibitem{HL-review2}
A.~Wang, 
``Ho\v{r}ava Gravity at a Lifshitz Point: A Progress Report", 
Int. J. Mod. Phys. D {\bf 26}, 1730014 (2017) [arXiv:1701.06087 [gr-qc]].

 
 \bibitem{power-counting_renormalizability}
T.~Fujimori, T.~Inami, K.~Izumi and T.~Kitamura,
``Power-counting and renormalizability in Lifshitz scalar theory", 
Phys. Rev. D {\bf 91}, 125007 (2015) [arXiv:1502.01820[hep-th]];
T.~Fujimori, T.~Inami, K.~Izumi and T.~Kitamura,
``Tree-Level Unitarity and Renormalizability in Lifshitz Scalar Theory", 
Prog. Theor. Exp. Phys. {\bf 2016}, 013B08 (2016) [arXiv:1510.07237[hep-th]].



\bibitem{ref_renormalize_HL}
 A.~O.~Barvinsky, D.~Blas, M.~Herrero-Valea, S.~M.~Sibiryakov and  C.~F.~Steinwachs,
 ``Renormalization of Horava gravity", 
 Phys. Rev. D {\bf 93}, 064022 (2016) [arXiv:1512.02250[hep-th]].


\bibitem{HL-review1}
S.~Mukhoyama,
``Ho\v{r}ava-Lifshitz cosmology: a review", 
Class. Quant. Grav. {\bf 27}, 223101 (2010) [arXiv:1007.5199[hep-th]].



\bibitem{CP1}
S.~Mukhoyama,
 ``Scale-invariant cosmological perturbations from Ho\v{r}ava-Lifshitz gravity without inflation", 
JCAP {\bf 090}, 001 (2009) [arXiv:0904.2190[hep-th]].

\bibitem{CP2}
 R.~Cai, B.~Hu and H.~Zhang, 
 ``Dynamical Scalar Degree of Freedom in Horava-Lifshitz Gravity", 
Phys. Rev. D {\bf 80}, 041501 (2009) [arXiv:0905.0255[hep-th]].

\bibitem{CP3}
T.~Kobayashi, Y.~Urakawa and M.~Yamaguchi, 
``Large scale evolution of the curvature perturbation in Horava-Lifshitz cosmology", 
JCAP {\bf 0911}, 015 (2009) [arXiv:0908.1005[astro-ph.CO]].

\bibitem{CP4}
A.~Wang, D.~Wands and R.~Maartens,
 ``Scalar field perturbations in Ho\v{r}ava-Lifshitz cosmology", 
JCAP {\bf 1003} 013 (2010) [arXiv:0909.5167[hep-th]];
A.~Wang,
 ``Vector and tensor perturbations in Horava-Lifshitz cosmology", 
 Phys. Rev. D {\bf 82}, 124063 (2010) [arXiv:1008.3637[hep-th]].

\bibitem{CP5}
K.~Izumi, T.~Kobayashi and S.~Mukohyama, 
``Non-Gaussianity from Lifshitz Scalar", 
JCAP {\bf 1010}, 031 (2010) [arXiv:1008.1406 [hep-th]].


\bibitem{HLC1}
T.~Takahashi, and J.~Soda,
 ``Chiral Primordial Gravitational Waves from a Lifshitz Point", 
Phys. Rev. Lett. {\bf 102}, 231301 (2009) [arXiv:0904.0554[hep-th]].


\bibitem{GW-HL}
A.~E.~G\"{u}mr\"{u}k\c{c}\"{u}o\u{g}lu, M.~Saravani and T.~P.~Sotiriou,
``Ho\v{r}ava Gravity after GW170817", 
Phys. Rev. D {\bf 97}, 024032 (2018) [arXiv:1711.08845[gr-qc]].



\bibitem{HLC2}
S.~Mukohyama, K.~Nakayama, F.~Takahashi, S.~Yokoyama,
 ``Phenomenological aspects of Ho\v{r}ava-Lifshitz cosmology", 
Phys .Lett. B {\bf 679}, 6-9 (2009) [arXiv:0905.0055[hep-th]].


 \bibitem{DM_integrate}
 S.~Mukohyama, 
 ``Dark matter as integration constant in Horava-Lifshitz gravity", 
 Phys. Rev. D {\bf 80}, 064005 (2009) [arXiv:0905.3563[hep-th]].

\bibitem{HLC3}
A.~Wang and Y.~Wu, 
``Thermodynamics and classification of cosmological models in the Horava-Lifshitz theory of gravity"
JCAP {\bf 0907} 012 (2009) [arXiv:0905.4117[hep-th]].

\bibitem{dS_projectable}
Y.~Q.~Huang, A.~Wang and Q.~Wu, 
``Stability of the de Sitter spacetime in Horava-Lifshitz theory",
Mod. Phys. Lett. {\bf A25} 2267 (2010) [arXiv:1003.2003[hep-th]];
A.~Wang and Q.~Wu,
``Stability of spin-0 graviton and strong coupling in Horava-Lifshitz theory of gravity",
Phys.~Rev.~D {\bf 83} 044025 (2011) [arXiv:1009.0268[hep-th]].

\bibitem{bounce_HL1}
G.~Calcagni,
 ``Cosmology of the Lifshitz universe", 
JHEP {\bf 0909}, 112 (2009) [arXiv:0904.0829[hep-th]].
\bibitem{bounce_HL2} 
E.~Kiritsis and G.~Kofinas, 
``Horava-Lifshitz Cosmology", 
Nucl. Phys. B{\bf 821}: 467-480 (2009) [arXiv:0904.1334[hep-th]].
\bibitem{bounce_HL3}
R.~H.~Brandenberger, 
``Matter Bounce in Horava-Lifshitz Cosmology", 
Phys. Rev. D{\bf 80}, 043516 (2009) [arXiv:0904.2835[hep-th]].
\bibitem{bounce_HL4}
Y.~F.~Cai, and E.~N.~Saridakis,
``Non-singular cosmology in a model of non-relativistic gravity ", 
JCAP {\bf 0910} 020 (2009) [arXiv:0906.1789[hep-th]].

\bibitem{previous_oscillating_universe}
K.~Maeda, Y.~Misonoh and T.~Kobayashi, 
``Oscillating Universe in Horava-Lifshitz Gravity"
Phys. Rev. {\bf D 82}, 064024 (2010) [arXiv:1006.2739[hep-th]].

\bibitem{Bianchi_IX}
Y.~Misonoh, K.~Maeda and T.~Kobayashi, 
``Oscillating Bianchi IX Universe in Horava-Lifshitz Gravity"
Phys. Rev. {\bf D 84}, 064030 (2011) [arXiv:1104.3978[hep-th]].

\bibitem{previous_stability_bounce}
Y.~Misonoh, M.~Fukushima, and S.~Miyashita, 
``Stability of singularity-free cosmological solutions in Ho\v{r}ava-Lifshitz gravity", 
Phys. Rev. D {\bf 95}, 044044 (2017) [arXiv:1612.09077[gr-qc]].

\bibitem{problem_projectable1}
C.~Charmousis, G.~Niz, A.~Padilla and P.~M.~Saffin,
 ``Strong coupling in Ho\v{r}ava gravity", 
JHEP {\bf 0908}, 070 (2009) [arXiv:0905.2579[hep-th]].

\bibitem{problem_projectable2}
M.~Li and Y.~Pang,
``A trouble with Ho\v{r}ava-Lifshitz gravity", 
JHEP {\bf 0908}, 015 (2009) [arXiv:0905.2751[hep-th]].

\bibitem{problem_projectable3}
D.~Blas, O.~Pujolas and S.~Sibiryakov,
``On the extra mode and inconsistency of Ho\v{r}ava gravity", 
JHEP {\bf 0910}, 029 (2009) [arXiv:0906.3046[hep-th]].

\bibitem{problem_projectable4}
K.~Koyama and F.~Arroja,
 ``Pathological behaviour of the scalar graviton in Ho\v{r}ava-Lifshitz gravity", 
JHEP {\bf 1003}, 061 (2010) [arXiv:0910.1998[hep-th]].

\bibitem{healthy_extension}
D.~Blas, O.~Pujolas, S.~Sibiryakov, 
``Consistent Extension of Ho\v{r}ava Gravity", 
Phys. Rev. Lett. {\bf 104}, 181302 (2010) [arXiv:0909.3525[hep-th]];
D.~Blas, O.~Pujolas, S.~Sibiryakov, 
``Comment on “Strong coupling in extended Ho\v{r}ava-Lifshitz gravity” [Phys. Lett. B 685 (2010) 197]", 
Phys. Lett. B {\bf 688}, 350 (2010) [arXiv:0912.0550[hep-th]];
D.~Blas, O.~Pujolas, S.~Sibiryakov, 
``Models of non-relativistic quantum gravity: the good, the bad and the healthy", 
JHEP {\bf 1104}, 018 (2011) [arXiv:1007.3503[hep-th]].

\bibitem{NP_stability1}
T.~Kobayashi, Y.~Urakawa, and M.~Yamaguchi,
``Cosmological perturbations in a healthy extension of Ho\v{r}ava gravity", 
JCAP {\bf 1004}, 025 (2010) [arXiv:1002.3101[hep-th]].

\bibitem{NP_stability2}
R.~G.~Cai, B.~Hu, and H.~B.~Zhang,
``Scalar graviton in the healthy extension of Ho\v{r}ava-Lifshitz theory", 
Phys. Rev. D {\bf 83}, 084009 (2011) [arXiv:1008.5048[hep-th]].

\bibitem{EA}
T.~Jacobson,
 ``Extended Ho\v{r}ava gravity and einstein-aether theory",
Phys. Rev. D {\bf 81}, 101502 (2010) [Phys. Rev. D {\bf 82}, 129901 (2010)] [arXiv:1001.4823[hep-th]].

 \bibitem{HL_BHTS1}
 D.~Blas and S.~Sibiryakov, 
 ``Ho\v{r}ava gravity versus thermodynamics: The black hole case”, 
 Phys. Rev. D {\bf 84}, 124043 (2011) [arXiv:1110.2195[hep-th]].

 \bibitem{HL_BHTS2}
 E.~Barausse, T.~Jacobson and T.~P.~Sotiriou, 
 ``Black holes in Einstein-aether and Ho\v{r}ava-Lifshitz gravity”, 
 Phys. Rev. D {\bf 83}, 124043 (2011) [arXiv:1104.2889[gr-qc]].

 \bibitem{HL_BHTS3}
 P.~Berglund, J.~Bhattacharyya and D.~Mattingly, 
 ``Mechanics of universal horizons", 
 Phys. Rev. D {\bf 85}, 124019 (2012) [arXiv:1202.4497[hep-th]].

\bibitem{HL_BHTS4}
K.~Lin, E.~Abdalla, R.~G.~Cai, A.~Wang,
 ``Universal horizons and black holes in gravitational theories with broken Lorentz symmetry", 
Int. J. Mod. Phys. D {\bf 23}, 1443004 (2014) [arXiv:1408.5976[gr-qc]].

\bibitem{HL_BHTS5}
K.~Lin, V.~H.~Satheeshkumar, A.~Wang,
 ``Static and rotating universal horizons and black holes in gravitational theories with broken Lorentz invariance", 
Phys. Rev. D {\bf 93}, 124025 (2016) [arXiv:1603.05708[gr-qc]].

 \bibitem{HL_BHTS6}
 Y.~Misonoh and K. Maeda, 
 ``Black Holes and Thunderbolt Singularities with Lifshitz Scaling Terms", 
 Phys. Rev. D {\bf 92}, 084049 (2015) [arXiv:1509.01378[gr-qc]].

\bibitem{inflation_Horava}
E.~G.~M.~Ferreira and R.~Brandenberger, 
``Trans-Planckian problem in the healthy extension of Ho\v{r}ava-Lifshitz gravity", 
Phys. Rev. D {\bf 86}, 043514 (2012) [arXiv:1204.5239[hep-th]].

\bibitem{SVW_paper}
T.~P.~Sotiriou, M.~Visser, S.~Weinfurtner, 
``Phenomenologically viable Lorentz-violating quantum gravity", 
Phys.Rev. Lett. {\bf 102}, 251601 (2009) [arXiv:0904.4464[hep-th]]; 
T.~P.~Sotiriou, M.~Visser, S.~Weinfurtner, 
``Quantum gravity without Lorentz invariance", 
JHEP {\bf 0910}, 033 (2009) [arXiv:0905.2798[hep-th]].

\bibitem{GW}
B.~P.~Abbott et al. [LIGO Scientific Collaboration, Virgo Collaboration], 
``GW170817: Observation of Gravitational Waves from a Binary Neutron Star Inspiral", 
Phys. Rev. Lett. {\bf 119}, 161101 (2017) [arXiv:1710.05832[gr-qc]];
B.~P.~Abbott et al. [LIGO Scientific Collaboration, Virgo Collaboration, Fermi Gamma-Ray Burst Monitor, INTEGRAL], 
``Gravitational Waves and Gamma-rays from a Binary Neutron Star Merger: GW170817 and GRB 170817A", 
Astrophys. J. {\bf 848} no.2, L13 (2017) [arXiv:1710.05834 [astro-ph.HE]].

\bibitem{ref_Hill_eq}
G.~W.~Hill, 
``On the part of the motion of the lunar perigee which is a function of the mean motions of the sun and moon", 
Acta Math. {\bf8}, 1 (1886).

\bibitem{U1_PHL1}
P.~Horava, and C.~M.~Melby-Thompson, 
``General Covariance in Quantum Gravity at a Lifshitz Point", 
Phys.Rev. D {\bf 82}, 064027 (2010) [arXiv:1007.2410[hep-th]].

\bibitem{U1_PHL2}
A.~Wang, and Y.~Wu, 
``Cosmology in nonrelativistic general covariant theory of gravity", 
Phys.Rev. D {\bf 83}, 044031 (2011) [arXiv:1009.2089[hep-th]].

\bibitem{U1_NPHL1}
T.~Zhu, Q.~Wu, A.~Wang, and F.~Shu,
``U(1) symmetry and elimination of spin-0 gravitons in Horava-Lifshitz gravity without the projectability condition", 
Phys. Rev. D {\bf 84}, 101502 (2011) [arXiv:1108.1237[hep-th]].

\bibitem{U1_NPHL2}
T.~Zhu, F.-W.~Shu, Q.~Wu, and A.~Wang,
``General covariant Horava-Lifshitz gravity without projectability condition and its applications to cosmology", 
Phys. Rev. D {\bf 85}, 044053 (2012) [arXiv:1110.5106[hep-th]].


 \end{thebibliography}
\end{document}